\RequirePackage{ifpdf}
\ifpdf 
\documentclass[pdftex]{sigma}
\else
\documentclass{sigma}
\fi

\let\t\tilde
\let\mc\mathcal

\let\bsy\boldsymbol

\let\vph\varphi
\let\bb\mathbb
\let\ge\geqslant
\let\le\leqslant
\def\p{\partial}
\newcommand{\ord}{\mathop{\rm ord}\nolimits}
\newcommand{\res}{\mathop{\rm res}\nolimits}
\newcommand{\Tr}{\mathop{\rm Tr}\nolimits}
\newcommand{\cond}[1]{\rule[-.8em]{.4pt}{1.6em}_{\ #1}}
\newcommand{\I}{\mathop{\rm Im}\nolimits}

\begin{document}

\allowdisplaybreaks

\numberwithin{equation}{section}

\renewcommand{\PaperNumber}{018}

\FirstPageHeading

\ShortArticleName{Integrable Evolutionary Systems and Their
Dif\/ferential Substitutions}

\ArticleName{Two-Field Integrable Evolutionary Systems of\\ the
Third Order  and Their Dif\/ferential Substitutions}

\Author{Anatoly G. MESHKOV and Maxim Ju. BALAKHNEV}

\AuthorNameForHeading{A.G. Meshkov and M.Ju. Balakhnev}

\Address{Orel State Technical University, Orel, Russia}
\Email{\href{mailto:a_meshkov@orel.ru}{a\_meshkov@orel.ru},
\href{mailto:maxibal@yandex.ru}{maxibal@yandex.ru}}

\ArticleDates{Received October 04, 2007, in f\/inal form January
17, 2008; Published online February 09, 2008}

\Abstract{A list of forty third-order exactly integrable
two-f\/ield evolutionary systems is presented. Dif\/ferential
substitutions connecting various systems from the list are found.
It is proved that all the systems can be obtained from only two of
them. Examples of zero curvature representations with $4\times4$
matrices are presented.}

\Keywords{integrability; symmetry; conservation law;
dif\/ferential substitutions; zero curvature representation}
\Classification{37K10; 35Q53; 37K20}

\section{Introduction}\label{introduction}
 We use the term ``integrability'' in the meaning that a system or equation under consideration possesses a Lax representation or a zero
curvature representation. Such systems can be solved by the
inverse spectral transform method (IST) \cite{Ab,Z}. Exactly
integrable evolution systems are of interest both for mathematics
and applications. In particular, systems of the following form
\begin{gather}
 u_t=u_{xxx}+F(u,v,u_x,v_x,u_{xx},v_{xx}),\qquad v_t=a\,v_{xxx}+G(u,v,u_x,v_x,u_{xx},v_{xx}), \label{sys0}
\end{gather}
where $a$ is a constant, excite great interest since about 1980.
The paper  \cite{DS1} is devoted to construction of systems of the
form (\ref{sys0}) among others. Nine integrable systems of the
form (\ref{sys0}) and their Lax representations have been obtained
in the paper. In particular, it contains a complete list of three
integrable systems  (\ref{sys0}) satisfying the conditions
$a(a-1)\ne0$ and $\ord(F,G)<2$. Here\, $\ord $ = order, $\ord f<n$
means that $f$ does not depend on
$u_{n},v_n,u_{n+1},v_{n+1},\dots$. Here and in what follows, the
notations  $u_n=\p^n u/\p x^n$, $v_n=\p^n v/\p x^n$ are used.

Two of the three mentioned systems can be written in the following
form
\begin{alignat}{3}\label{s1}
&  u_t =u_3+v\, u_1, \qquad && v_t=-\tfrac 1 2 v_3+u\,u_1-v\,v_1,& \\
&  u_t=u_3+v\,u_1, \qquad && v_t=-\tfrac 1 2 v_3-v\,v_1+u_1
\label{s2}&
\end{alignat}
and the third system is presented below (see (\ref{ex5})). System
(\ref{s1}) was found independently  in \cite{HS} and the soliton
solutions were constructed there. This system is called as the
Drinfeld--Sokolov--Hirota--Satsuma system.

This paper contains two results: (i) a list of integrable systems
of the form (\ref{sys0}) with smooth functions $F$,~$G$ and
$a=-1/2$; (ii) dif\/ferential substitutions that allow to connect
any equation from the list with (\ref{s1}) or (\ref{s2}).

There are many articles dealing with integrable systems,  but some
of them (see, e.g., \cite{AF,Ma,MaP}) consider multi-component
systems. Other papers (see, e.g., \cite{Fuch,Nu,G1,Fo1,Zh})
contain two-component systems reducible to a triangular form. The
triangular form is brief\/ly considered below. There was possibly
only one serious attempt~\cite{Ka} to classify integrable systems
of the form~(\ref{sys0}) using the Painlev\'e test. Unfortunately,
f\/ifteen systems presented in \cite{Ka} contain a~large number of
constants some of which can be removed by scaling and linear
transformations. Note that there are two-f\/ield integrable
evolutionary systems $\bsy u_t=A\bsy u_3+\bsy H(\bsy u,\bsy
u_1,\bsy u_2)$ with a non-diagonal main matrix $A$. For example,
an integrable evolutionary system with the Jordan main matrix is
found in \cite{Fo}.

Moreover, about 50 two-f\/ield integrable systems of the form
(\ref{sys0}) with $a=1$ can be extracted from papers
\cite{Sph,SokMesh,Bal,BalMesh,BalM} that deal with vector
evolutionary equations.

Partial solutions of the classif\/ication problem for $a=0$ and
$\ord G\le1$ have been obtained in~\cite{Ku}, and in \cite{FPM}
for divergent systems with $a\ne1$. A complete list of integrable
systems of the form (\ref{sys0}) does not exist today because the
problem is too cumbersome and the set of integrable systems is
very large.

Our tool is the symmetry method presented in many papers. We shall
point out pioneer or review papers only. In~\cite{Sh1} the notions
of formal symmetry and canonical conserved density for a~scalar
evolution equation are introduced. These tools were applied to
classif\/ication of the KdV-type equations in \cite{svsok1}. A
complete theory of formal symmetries and formal conservation laws
for scalar equations has been presented in~\cite{Sh3}.  A
generalized theory was developed for evolutionary systems
in~\cite{MSY}. Review paper~\cite{MSS} contains both general
theorems of the symmetry method and classif\/ication results on
integrable equations: the third and f\/ifth order scalar
equations, Schr\"odinger-type systems, Burgers-type equations and
systems. Review paper \cite{fokas} is devoted to higher
symmetries, exact integrability and related problems.
Peculiarities of systems~(\ref{sys0}) have been discussed
in~\cite{FPM}. For the sake of completeness, the main points of
the symmetry method and some results necessary for understanding
of this paper are considered in the Sections
\ref{sec2}--\ref{sec4}.

Brief\/ly speaking, the symmetry  method deals with the so-called
canonical conservation laws
\begin{gather}\label{law}
D_t\rho_n=D_x\theta_n,\qquad D_t\t\rho_n=D_x\t\theta_n,\qquad
n=0,1,2,\dots,
\end{gather}
where $D_t$ is the evolutionary derivative and $D_x$ is the total
derivative with respect to $x$. In particular,
$\rho_0=-F_{u_2}/3$, $ \t\rho_0=-G_{v_2}/(3a)$. The recursion
relations for the canonical conserved  densities $\rho_n$ and
$\t\rho_n$ are presented in Section~\ref{sec2}. All canonical
conserved densities are expressed in terms of functions $F$ and
$G$. That is why equations (\ref{law}) impose great restrictions
on the forms of $F$ and $G$. Equations (\ref{law}) are solvable in
the jet space if\/f
\begin{gather}\label{con1}
E_\alpha D_t\rho_n=0,\qquad E_\alpha D_t\t\rho_n=0,\qquad \alpha
=1,2,\qquad n=0,1,2,\dots
\end{gather}
(see \cite{G}, for example). Here
\[
E_\alpha\equiv \frac{\delta }{\delta u^\alpha }=\sum_{n=0}^{\infty
} (-D_x)^n\frac{\p }{\p u_n^\alpha },\qquad (u^1=u,\ u^2=v),
\]
is the Euler operator.

 Conservation law with $\rho =D_x\chi,\theta =D_t\chi $ is called trivial and the conserved density of the form $\rho =D_x\chi$ is called trivial too.
This can be written in the form $\rho\in \I D_x$, where $\rm
Im=Image$. If $\rho_1-\rho_2\in \I D_x$, then the densities
$\rho_1$ and $\rho_2$ are said to be equivalent.

There are a lot of systems in the following form
\begin{gather*}
u_t=u_{xxx}+F(u,u_x,u_{xx}),\qquad v_t=a\,v_{xxx}+G(u,v,u_x,v_x,u_{xx},v_{xx}), 
\end{gather*}
satisfying the  integrability conditions~(\ref{law}). Such systems
containing one independent equation are said to be triangular. It
follows from the integrability conditions that the equation for
$u$ must be one of the known integrable equations (KdV, mKdV etc).
The second equation is usually linear with respect to $v$, $v_x$
and $v_{xx}$. Triangular systems do not possess any Lax
representations and are not integrable in this sense. Therefore
triangular systems and those reducible to the triangular form have
been omitted as trivial.

The system of two independent equations
\begin{gather*}
u_t=u_{xxx}+F(u,u_x,u_{xx}),\qquad v_t=a\,v_{xxx}+G(v,v_x,v_{xx}), 
\end{gather*}
will be called disintegrated. It is obvious that the disintegrated
form is a partial case of the triangular form. Therefore the
disintegrated systems and those reducible to them have been
omitted.

System (\ref{sys0}) will be called reducible if it is triangular
or can be reduced to triangular or disintegrated form. Otherwise,
the system will be called irreducible.

Our computations show that for irreducible integrable systems
(\ref{sys0})  parameter $a$ must belong to the following set:
\[
A=\left\{0,\ -2,\ -\tfrac12,\  -\tfrac72+\tfrac32\sqrt{5}, \
-\tfrac72-\tfrac32\sqrt{5}\right\}.
\]
These values were found f\/irst in \cite{DS1} and were repeated in
\cite{D-S}. The value of $a$ is always def\/ined at the end of
computations when functions $F$ and $G$ have been found and only
some coef\/f\/icients are to be specif\/ied from the fifth or
seventh integrability conditions (see example in
Section~\ref{sec3.1}). This means that it is enough to verify
conditions (\ref{con1}) for $n=0,\dots,7$ and $\alpha =1,2$ to
obtain~$F$,~$G$ and~$a$. But for absolute certainty we have
verif\/ied  conditions (\ref{con1}) for $n=8,9$ and $\alpha =1,2$
for each system.

The presented set $A$ consists of zero and two pairs $(a,a^{-1})$.
The transformation $t'=at$, $u'=v$, $v'=u$ changes the parameter
$a\ne0$ in (\ref{sys0}) into $a^{-1}$. That is why one ought to
consider the values $\left\{0,\;  -\frac{1}{2},\;
-\frac{7}{2}+\frac{3}{2}\sqrt{5}\right\}$ of  the parameter $a$.
Integrable systems with $a=0$ were mentioned above, see
also~\cite{Max}. This paper is devoted to investigation of the
case $a=-1/2$ only. The case $a=-\frac72+\frac32\sqrt{5}$ will be
presented in another paper.

Section~\ref{sec2} contains recursion formulas for the canonical
densities. The origin of the notion, some examples and a
preliminary classif\/ication are considered.

A list of forty integrable systems and an example of computations
are presented in Section~\ref{sec3}.

Section \ref{sec4} contains dif\/ferential substitutions that
connect all systems from the list. The method of computations and
an example are considered. It is shown that all systems from the
list presented in Section \ref{sec3} can be obtained
from~(\ref{s1}) and~(\ref{s2}) by dif\/ferential substitutions.

Section \ref{sec5} is devoted to zero curvature representations.
The zero curvature representations for systems~(\ref{s1}) and
(\ref{s2}) are obtained from the Drinfeld--Sokolov $L$-operators.
A method of obtaining zero curvature representations for other
systems is demonstrated.

\section{Canonical densities}\label{sec2}

One of the main objects of the symmetry approach to
classif\/ication of integrable equations is the inf\/inite set of
the canonical conserved densities. Let us demonstrate how
canonical conserved densities can be obtained the from the
asymptotic expansions for eigenfunctions of the Lax operators. The
simplest Lax equations concerned with the KdV equation
\[
u_t=6uu_x-u_{xxx}
\]
take the following form
\begin{gather}
\psi _{xx}-u\psi -\mu^2\psi =0,\label{lax1}\\
\psi_t=-4\psi_{xxx}+6u\psi_x+3u_x\psi +4\mu^3\psi .\label{lax2}
\end{gather}
Here $u$ is a solution of the  KdV equation and $\mu$ is a
parameter. The standard substitution
\[
\psi =\exp\left(\int\rho\,dx\right)
\]
reduces equations (\ref{lax1}) and  (\ref{lax2}) to the Riccati
form
\begin{gather}
\rho_x+\rho^2-u-\mu^2=0, \label{ric}\\
\p_t\int\rho\,dx=-4(\p_x+\rho)^2\rho +
6u\rho+3u_x+4\mu^3.\label{tmp}
\end{gather}
Dif\/ferentiating temporal equation (\ref{tmp}) with respect to
$x$ one can rewrite it, using (\ref{ric}), as the continuity
equation:
\begin{gather}\label{seql}
\rho_t=\p_x[(2u-4\mu^2)\rho-u_x].
\end{gather}
To construct an asymptotic expansion one ought to set
\begin{gather}\label{exp}
\rho=\mu+\sum_{n=0}^{\infty } \rho_n(-2\mu)^{-n}.
\end{gather}
Then equation (\ref{ric}) results in the following well known
recursion formula \cite{Z}
\begin{gather}\label{den}
\rho_{n+1}=D_x\rho_n+\sum_{i=1}^{n-1} \rho_i\rho_{n-i},\qquad
n=1,2,\dots,\qquad \rho_0=0,\qquad \rho_1=- u,
\end{gather}
and  (\ref{seql}) results in inf\/inite sequence of conservation
laws:
\begin{gather}\label{ll}
D_t\rho_n=D_x(2u\rho_n-\rho_{n+2}),\qquad n>0.
\end{gather}
We change here $\p_t\to D_t$ and  $\p_x\to D_x$ because $u$ is a
solution of the KdV equation. The obtained conservation laws are
canonical. It is easy to obtain several f\/irst canonical
densities:
\[
\rho_2=- u_1,\qquad \rho_3=u^2-u_2,\qquad
\rho_4=D_x(2u^2-u_2),\qquad \dots.
\]
It  is shown in \cite{Z} that all even canonical densities are
trivial. Note that if one chooses another asymptotic expansion,
for example, in powers of $\mu^{-1}$ instead of (\ref{exp}), then
another set of canonical densities is obtained, which is
equivalent to the previous set.

The canonical densities that follow from (\ref{den})  can also be
obtained by using the temporal equation (\ref{tmp}) only. Indeed,
setting $\p_t\int\rho\,dx=\theta$ one obtains from (\ref{tmp})
\begin{gather}\label{pro}
-4(\p_x+\rho)^2\rho +6u\rho+3u_x+4\mu^3=\theta.
\end{gather}
Using the same expansions as above
\[
\rho=\mu+\sum_{n=0}^{\infty } \rho_n(-2\mu)^{-n},\qquad
\theta=\sum_{n=0}^{\infty } \theta_n(-2\mu)^{-n},
\]
one can obtain from (\ref{pro}) the following recursion relation:
\begin{gather*}
\rho_{n+2} =2u\rho_{n}+2\sum_{i=0}^{n+1} \rho_{i}\rho_{n-i+1}-\tfrac{4}{3}\sum_{i,j=0}^{n}\rho_{i}\rho_{j}\rho_{n-i-j}-\tfrac13\theta_n\\
\phantom{\rho_{n+2} =}{} +2D_x\left(\rho_{n+1} -\sum_{i=0}^{n}
\rho_{i}\rho_{n-i}\right)-\tfrac43 D_x^2\rho_{n} -u\delta
_{n,-1}+u_1\delta_{n0}, \qquad n=-2,-1,0,\dots,
\end{gather*}
where  $\delta_{i,k}$ is the Kronecker delta. The obtained
relation provides $\rho_0=0$, $\rho_1=-u$,
$\rho_2=-u_1-\theta_0/3$, etc. As $D_t\rho_0=D_x\theta_0$ and
$\rho_0=0$, then $\theta_0=0$. The higher canonical densities
$\rho_n$, $n>2$ depend on $\theta_{n-2}$. The f\/luxes $\theta_n$
must be def\/ined now from equations (\ref{law}). For example,
$\theta_1=u_2-3u^2$.

The traditional method to obtain the  canonical densities for an
evolution system \cite{MSY}
\begin{gather}\label{SYS}
\bsy u_t=\bsy K(\bsy u,\bsy u_x,\dots,u_n), \qquad \bsy u(t,x)\in
\bb R^m,\quad m\ge1,\qquad u^\alpha_k=\p_x^k u^\alpha.
\end{gather}
consists, brief\/ly, in the following. The main idea is to use the
linearized equation
\begin{gather}\label{LIN}
(D_t- \bsy K_*)\psi =0
\end{gather}
or its adjoint
\begin{gather}\label{aLIN}
(D_t+ \bsy K_*^+)\vph =0
\end{gather}
as the temporal Lax equation. Here
\begin{gather*}
(\bsy K_*\psi)^\alpha =\sum_{n,\beta } \frac{\p K^\alpha }{\p
u^\beta _n } D_x^n\psi ^\beta ,\qquad
(\bsy K_*^+\vph)_\alpha =\sum_{n,\beta }  (-D_x)^n\frac{\p K^\beta  }{\p u^\alpha _n }\vph_\beta, \\
D_t=\frac{\p}{\p t}+\sum_{n,\alpha } D_x^n(K^\alpha )\frac{\p}{\p
u^\alpha_n},\qquad D_x=\frac{\p}{\p x}+\sum_{n,\alpha }
u^\alpha_{n+1}\frac{\p}{\p u^\alpha_n}.
\end{gather*}
The spatial Lax operator (formal symmetry) was introduced in
\cite{MSY} as the inf\/inite operator series
\begin{gather}\label{rec}
R=\sum_{k=-\infty }^{N} R_kD_x^k,\qquad N>0,
\end{gather}
 commuting with $D_t- \bsy K_*$. $R_k$ are matrix coef\/f\/icients depending on $\bsy u,\bsy u_x,\dots $. It was shown that
$\Tr\,\res R$ ($\res R= R_{-1}$) is the conserved density for
system (\ref{SYS}). Canonical densities have been def\/ined by the
formulas
\[
\rho_n=\Tr\,\res R^n,\qquad n=1,2,\dots,
\]
see \cite{MSS} for details.

Operations with operator series (\ref{rec}) are not so simple,
therefore we use an alternative method for obtaining the
canonical densities. It was proposed in \cite{Ch} heuristically
and we present the following explanation (see also \cite{IP}).

{\bf Observation.} One can obtain equation (\ref{pro}) from
(\ref{lax2}) by the following substitution
\begin{gather}\label{subs}
\psi =e^\omega ,\qquad \omega = \int \rho\,dx+\theta \,dt,
\end{gather}
where $\rho\,dx+\theta \,dt$ is the smooth closed 1-form,
that is, $D_t\rho =D_x\theta $. This implies $e^{-\omega }D_t
e^\omega =D_t+\theta $, $e^{-\omega }D_x e^\omega =D_x+\rho$ and
so (\ref{pro}) follows. Another way to obtain the same equation is
to prolong the operators $D_t \to\p_t+\theta$, $D_x\to\p_x+\rho$
in (\ref{lax2}) formally and to set $\psi =1$. For systems, one
must set $\psi _\alpha =1$ for a f\/ixed $\alpha $ only.

We shall apply this method to system (\ref{sys0}) now.

The linearized system (\ref{sys0}) with prolonged operators
$D_x\to D_x+\rho$, $D_t\to D_t+\theta$ takes the following form:
\begin{gather}
\big[(D_x+\rho)^3+F_u+F_{u_1}(D_x+\rho)+F_{u_2}(D_x+\rho)^2-D_t-\theta \big]\Psi_1\nonumber\\
\qquad{}+\big[ F_v+F_{v_1}(D_x+\rho)+F_{v_2}(D_x+\rho)^2\big]\Psi_2=0,  \nonumber\\
\big[ G_u+G_{u_1}(D_x+\rho)+G_{u_2}(D_x+\rho)^2\big]\Psi_1+ a\,(D_x+\rho)^3\Psi_2\nonumber\\
\qquad{}+\big[G_v+G_{v_1}(D_x+\rho)+G_{v_2}(D_x+\rho)^2-D_t-\theta\big]\Psi_2=0.\label{lin}
\end{gather}
If one sets here $\Psi_1=1$, then the f\/irst equation takes the
following form
\begin{gather*}
(D_x+\rho)^2\rho +F_u+F_{u_1}\rho+F_{u_2}(D_x+\rho)\rho -\theta \\
\qquad{}+\big[
F_v+F_{v_1}(D_x+\rho)+F_{v_2}(D_x+\rho)^2\big]\Psi_2=0.
\end{gather*}
It is obvious from this equation that the following forms of the
asymptotic expansions are acceptable:
\[
\rho=\mu^{-1}+\sum_{n=0}^{\infty }\rho_n\,\mu^n,\qquad \theta
=\mu^{-3}+\sum_{n=0}^{\infty }\theta _n\,\mu^n, \qquad
\Psi_2=\sum_{n=0}^{\infty } \rho_n\,\mu^n.
\]
Here $\mu$ is a complex parameter. Then, after some simple
calculations, the following recursion relations are obtained
$(n\ge -1)$:
\begin{gather*}
\rho_{n+2} =\tfrac1 3\theta_n-\sum_{i=0}^{n+1} \rho_i \rho_{n-i+1}
-\tfrac1 3\sum_{i+j=0}^{n}\rho_i \rho_j \rho_{n-i-j}
-\tfrac1 3\,F_{u_1}(\delta_{n,-1}+\rho_n)-\tfrac1 3\,F_u\,\delta_{n,0} \\
\phantom{\rho_{n+2} =}{} -\tfrac1
3(F_v+F_{v_1}D_x+F_{v_2}D_x^2)\vph_n
-\tfrac1 3\,F_{u_2}\left(D_x\rho_n+2\,\rho_{n+1}+\sum_{i=0}^{n} \rho_i \rho_{n-i}\right)\\
\phantom{\rho_{n+2} =}{}-\tfrac1
3\,F_{v_2}\left(\vph_{n+2}+2\sum_{i=0}^{n}\rho_i
\vph_{n-i+1}+\sum_{i+j=0}^{n}
 \rho_i \rho_j \vph_{n-i-j}\right)\\
\phantom{\rho_{n+2} =}{}-\tfrac1 3\,F_{v_2}\left(2\,D_x\,\vph_{n+1}+\sum_{i=0}^{n} \rho_i D_x\,\vph_{n-i}+D_x\sum_{i=0}^{n} \rho_i \vph_{n-i}\right)\\
\phantom{\rho_{n+2} =}{}-D_x\left[\rho_{n+1}+\tfrac1
3\,D_x\,\rho_n+\tfrac1 2\sum_{i=0}^{n}\rho_i \rho_{n-i}\right]
-\tfrac1 3\,F_{v_1}\left(\vph_{n+1}+\sum_{i=0}^{n} \rho_i
\vph_{n-i}\right),
\\
(1-a)\vph_{n+3}=G_u\delta_{n,0}+G_{u_2}(D_x\rho_n+2\,\rho_{n+1}+\sum_{i=0}^{n} \rho_i \rho_{n-i})+G_{u_1}(\delta_{n,-1}+\rho_n)\\
\phantom{(1-a)\vph_{n+3}=}{} -\sum_{i=0}^{n} \theta_i \vph_{n-i}+G_v\,\vph_n +G_{v_1}\left(D_x\vph_n+\vph_{n+1}+\sum_{i=0}^{n}\rho_i\vph_{n-i}\right)-D_t\,\vph_n \\
\phantom{(1-a)\vph_{n+3}=}{}+G_{v_2}\left(2\,D_x\vph_{n+1}+\sum_{i=0}^{n}\rho_i\,D_x\,\vph_{n-i}
+D_x\sum_{i=0}^{n}\rho_i\vph_{n-i}\right)\\
\phantom{(1-a)\vph_{n+3}=}{}+G_{v_2}\left(\vph_{n+2}+D_x^2\vph_n +2\sum_{i=0}^{n+1}\rho_i\vph_{n-i+1}+\sum_{i+j=0}^{n}\rho_i \rho_j \vph_{n-i-j}\right)\\
\phantom{(1-a)\vph_{n+3}=}{}+a\,D_x^3\vph_n+3\,a\,D_x^2\vph_{n+1}+6\,a\sum_{i=0}^{n+1}\rho_i\,D_x\,\vph_{n-i+1}+3\,a\sum_{i=0}^{n+2} \rho_i \vph_{n-i+2}\\
\phantom{(1-a)\vph_{n+3}=}{}+3\,a\,D_x\vph_{n+2}+3\,a\sum_{i+j=0}^{n}\rho_i \rho_j\,D_x\,\vph_{n-i-j}+3\,a\sum_{i=0}^{n}\vph_{n-i+1}\,D_x\,\rho_i\\
\phantom{(1-a)\vph_{n+3}=}{} +\tfrac3 2\,a\sum_{i+j=0}^{n}\vph_{n-i-j}\,D_x(\rho_i \rho_j )+3\,a\sum_{i+j=0}^{n+1}\rho_i \rho_j \vph_{n-i-j+1}\\
\phantom{(1-a)\vph_{n+3}=}{}+3\,a\,D_x\sum_{i=0}^{n}
\rho_i\vph_{n-i}+a\sum_{i=0}^{n}
\vph_{n-i}\,D_x^2\rho_i+a\sum_{i+j+k=0}^n \rho_i \rho_j
\rho_k\,\vph_{n-i-j-k} .
\end{gather*}
Here $\delta_{i,k}$ is the Kronecker delta, $F_{u_1}=\p F/\p u_1$
and so on. From the recursion relations it is obvious why the
value $a=1$ is singular. Some of initial elements of the sequence
$\{\rho_n,\vph_n\}$ read
\[
\rho_0=-\tfrac1 3\,F_{u_2},\qquad \vph_0=0,\qquad
\vph_1=\frac{1}{1-a}\,G_{u_2},
\]
others are introduced via the $\delta$-symbols.

If one sets in (\ref{lin}) $\Psi_2=1$ and $a\ne 0$, then one more
pair of recursion relations for $\{\t\rho_n,\t\vph_n\}$ is
obtained. These recursion relations give us any desired number of
canonical densities. As an example, we present here some more
canonical densities:
\begin{gather}
 \rho_0=-\frac{1}{3}\,F_{u_2},\qquad
\rho_1=\frac{1}{9}\,F_{u_2}^2-\frac{1}{3}\,F_{u_1}+\frac{1}{3\,b}\,F_{v_2}G_{u_2}+\frac{1}{3}\,D_x\,F_{u_2},
\nonumber\\
 \t\rho_0=-\frac{1}{3\,a}\,G_{v_2},\qquad \t\rho_1=\frac{1}{9\,a^2}\,G_{v_2}^2-\frac{1}{3\,a}\,G_{v_1}
-\frac{1}{3\,a\,b}\,F_{v_2}G_{u_2}+\frac{1}{3\,a}\,D_x\,G_{v_2},
\label{dens}
\end{gather}
where $b=a-1$.  The tilde denotes another sequence of canonical
densities. Further canonical  densities are too cumbersome,
therefore we do not present them here.

To simplify investigation of the integrability conditions, an
additional requirement is always imposed. This is the existence of
a formal conservation law \cite{MSY,MSS}. A formal conservation
law is an operator series $N$ in powers of $D_x^{-1}$. An equation
for  the formal conservation law can be written in the following
operator form
\begin{gather}\label{N}
(D_t-{K_*})\mc N=\mc N(D_t+{K_*}^+).
\end{gather}
The form of this equation coincides with the form of  the equation
for the Noether operator \cite{Fu}. That is a formal conservation
law may be called a formal Noether operator.

If $(D_t-K_*,L)$ is the Lax pair for an equation, then
$(D_t+{K_*}^+,L^+)$ is obviously the Lax pair for the same
equation. Hence, canonical densities obtained from (\ref{LIN})
must be equivalent to canonical densities obtained from
(\ref{aLIN}).

It was shown in \cite{FPM} that the f\/irst sequence of the
canonical densities $\rho_n$ for system (\ref{sys0}) obtained from
(\ref{LIN})  is equivalent to the f\/irst sequence of the
canonical densities $\tau_n$ obtained from (\ref{aLIN}) and
 the second sequence of the canonical densities $\t\rho_n$ is equivalent to the second sequence of the canonical densities $\t\tau_n$.
Hence, $\rho_n-\tau_n\in \I D_x$ and  $\t\rho_n-\t\tau_n\in \I
D_x$, or
\begin{gather}
 E_\alpha (\rho_n-\tau_n)=0,\qquad  E_\alpha (\t\rho_n-\t\tau_n)=0,\qquad \alpha =1,2,\qquad n=0,1,2,\dots. \label{law2}
\end{gather}
Equations (\ref{law}) (or (\ref{con1})) and (\ref{law2}) are said
to be the necessary conditions of integrability. We shall refer
to it simply as the integrability conditions for brevity.

Our computations have shown that
\begin{gather}\label{cd-acd}
\tau_0=-\rho_0,\qquad \t\tau_0=-\t\rho_0,\qquad
\tau_1=\rho_1,\qquad \t\tau_1=\t\rho_1.
\end{gather}
Other  ``adjoint''  canonical densities $\tau_i$ and $\t\tau_k$
essentially dif\/fer from the ``main''  canonical densities $\rho
_i$ and $\t\rho _k$. All canonical densities can be obtained
using the Maple routines {\sl cd}  and {\sl acd} from the package
JET  (see \cite{JetM}). These routines generate the ``main''  and
the ``adjoint''  canonical densities, correspondingly, for almost
any evolutionary system (an exclusion is the case of multiple
roots of the main matrix of the system under consideration).

Thus, according to (\ref{dens}) and (\ref{cd-acd}) we have
$F_{u_2}\in \I D_x$ and  $G_{v_2}\in \I D_x$ ($a\ne0$).  This
implies the following lemma.

\begin{lemma}\label{lemma1} System \eqref{sys0} with $a(a-1)\ne0$ satisfying the zeroth integrability conditions \eqref{law2}  reads{\samepage
\begin{gather}
 u_t=u_3-\frac{3}{2f}\,u_2\,D_x\,f +\frac{3}{4f}\,f_{u_1}u_2^2+F_1(u,v,u_1,v_1,v_2),\nonumber \\
 v_t=av_3-\frac{3a}{2g}\,v_2\,D_x\,g + \frac{3a}{4g}\,g_{v_1}v_2^2+G_1(u,v,u_1,v_1,u_2).\label{sy2}
 \end{gather}
where $\ord(f,g)\le1$.}
\end{lemma}

Indeed, one may set $F_{u_2}=-3/2 D_x\ln f$ and $G_{v_2}=- 3/2 a
D_x\ln g$, where $\ord(f,g)\le1$ because $\ord(F,G)\le2$. Then
equations (\ref{sy2}) follow.

From higher integrability conditions one more lemma follows.

\begin{lemma}\label{lemma2} Suppose  system \eqref{sy2} is irreducible and satisfies the following eight integrability conditions
$\rho_2-\tau_2\in \I D_x$, $\t\rho_2-\t\tau_2\in \I D_x$ and
$D_t\rho_n\in \I D_x$, $D_t\t\rho_n\in \I D_x$, where $n = 1, 3,
5$. Then the system must have the following form
\begin{gather}
u_t=u_3-\frac{3}{2f}\,u_2\,D_x\,f +\frac{3}{4f}\,f_{u_1}u_2^2+f_1\,v_2^2+f_2\,v_2+f_3, \nonumber\\
v_t=av_3-\frac{3a}{2g}\,v_2\,D_x\,g +
\frac{3a}{4g}\,g_{v_1}v_2^2+g_1\,u_2^2+g_2\,u_2+g_3, \qquad
a\ne0,\label{sy2a}
 \end{gather}
where $\ord(f,g,f_i,g_j)\le1$.
\end{lemma}

A scheme of the proof has been presented in \cite{FPM}.

\section{List of  integrable systems}\label{sec3}

As it is shown in Section \ref{sec2} the problem of the
classif\/ication of integrable systems (\ref{sys0}) is reduced to
investigation of system (\ref{sy2a}). That is why it is necessary
to start by investigating its symmetry properties.

\begin{lemma}\label{lemma3} System \eqref{sy2a} are invariant under any point transformation of the form
\begin{gather*}
(a)\ \ t'=\alpha^3 t+\beta,\qquad x'=\alpha  x+\gamma  t+\delta ,\qquad \alpha\ne0,  \qquad u'=u,\qquad v'=v, \\
(b)\ \ u'=h_1(u),\qquad v'=h_2(v),
\end{gather*}
and under the following permutation transformation
\begin{gather*}
(c)\ \ t'=at,\qquad u'=v, \qquad v'=u,
\end{gather*}
where $\alpha$, $\beta$, $\gamma$ and $\delta$ are constants,
$h_i$ are arbitrary smooth functions.
\end{lemma}

The classif\/ication of systems of type (\ref{sy2a}) has been
performed by modulo of the presented transformations.

Moreover, some systems (\ref{sy2a}) admit invertible contact
transformations. An ef\/fective tool for searching such contact
transformations is investigation of the canonical conserved
densities. For example, system (\ref{11,1-0-0}) from the next
section has the f\/irst canonical conserved density of the
following form:
\[
\rho_1 =\left(v_1-\tfrac23 ue^v\right)^2+2c_1^2e^{-2v}.
\]
It is obvious that the best variables for that system are
\[
U=e^{-v}\qquad \text{and}\qquad V=v_1-\tfrac23 ue^v.
\]
This is an invertible contact transformation. In terms of $U$ and
$V$ the system takes the  following simple form:
\begin{gather*}
U_t=D_x\left(U_2+\tfrac32 UV_1-\tfrac34UV^2+\tfrac12c_1^2U^3\right),\\
 V_t=\tfrac14 D_x(V^3-2V_2)-\tfrac32 c_1^2D_x(2UU_1+U^2V).
\end{gather*}
If $c_1\ne0$ this system can be reduced to (\ref{1d}) by scaling,
otherwise the system is triangular: the equation for $V$ will be
independent single mKdV. Moreover, the equation for $U$ becomes
linear. That is why  $c_1\ne0$ in  (\ref{11,1-0-0}).

Canonical densities for the triangular systems contain only one
highest order term in the second power as in the considered
example $\rho =V^2$ or $\rho=V_x^2+\cdots$, or
$\rho=V_{xx}^2+\cdots$ etc. Triangular systems and those reducible
to the triangular form have been omitted in the classif\/ication
process as trivial.

To classify integrable systems (\ref{sys0}) with $a(a-1)\ne0$ one
must solve a huge number of large overdetermined partial
dif\/ferential systems for eight unknown functions of four
variables. This work has required powerful computers and has taken
about six years. All the calculations have been performed in the
interactive mode of operation because automatic solving of large
systems of partial dif\/ferential equations is still impossible.
The package {\sc pdsolve} from the excellent system Maple makes
errors solving some single partial dif\/ferential equations. The
package {\sc diffalg} cannot operate with large systems because
its algorithms are too cumbersome. Thus, one has to solve
complicated problems in the interactive mode. Hence, to obtain a
true solution one must enter true data! Under such circumstances
errors are probable. The longer the computations the more probable
are errors.
 This is the reason why we cannot state with conf\/idence that all computations have been precise all these six years.
That is why the statement on completeness of the obtained set of
integrable systems is formulated as a hypothesis.

In this and in the following sections  $c$, $c_i$, $k$, $k_i$ are
arbitrary constants.

\medskip

\noindent {\bf Hypothesis.} {\it Suppose  system \eqref{sy2a} with
$a=-1/2$ is irreducible. If  the system has infinitely many
canonical conservation laws, then it can be reduced by an
appropriate point transformation to one of the following systems:}
\begin{gather}\label{9}
u_t =u_3+v\, u_1,\qquad v_t=-\tfrac 1 2 v_3+u\,u_1-v\,v_1;
\\
u_t =u_3+v_1\, u_1,\qquad v_t=-\tfrac 1 2 v_3+\tfrac12
(u^2-v_1^2);\label{9-0}
\\
\label{8} u_t =u_3+v\,u_1,\qquad v_t=-\tfrac 1 2 v_3-v\,v_1+u_1;
\\
u_t =u_3+v\,u_1+v_1\, u,\qquad v_t=-\tfrac 1 2
v_3-v\,v_1+u;\label{8-1}
\\
u_t =u_3+v_1\,u_1,\qquad v_t=-\tfrac 1 2 v_3-\tfrac12
v_1^2+u;\label{8-2}
\\
u_t =u_3+u\, u_1+v_1,\qquad v_t=-\tfrac 1 2
v_3+\tfrac{3}{2}u_1u_2-u\,v_1;\label{5}
\\
\label{5-1} u_t =u_3+v_2+k\,u_1,\qquad v_t=-\tfrac 1 2
v_3+\tfrac{3}{2}uu_2+\tfrac{3}{4}u_1^2+\tfrac{1}{3}u^3+k\left(
u^2-v_1 \right);
\\
\label{5-1-1} u_t
=u_3+\tfrac{3}{2}vv_2+\tfrac{3}{4}v_1^2+\tfrac{1}{3}v^3-k\left(
v^2+u_1\right),\qquad v_t=-\tfrac 1 2 v_3+u_2+k\,v_1;
\\ \label{1}
u_t =u_3-\tfrac3 2u_1 v_2 -\tfrac 3 4 u_1 v_1^2 +\tfrac 1 4 u_1^3
, \qquad v_t=-\tfrac 1 2 v_3+\tfrac 3 2 u_1u_2-\tfrac 3 4
u_1^2v_1+\tfrac 1 4 v_1^3;
\\
\label{1d} u_t
=\left(u_2-\tfrac32\,uv_1-\tfrac34\,uv^2+\tfrac14\,u^3\right)_x,\qquad
v_t=\left(-\tfrac12\,v_2+\tfrac32
uu_1-\tfrac34u^2v+\tfrac14v^3\right)_x;
\\
\label{9-40} u_t
=u_3-\tfrac{3}{2}\,v_2-\tfrac{3}{2}\,u_1v_1-\tfrac{1}{2}\,u_1^3,\qquad
v_t=-\tfrac 1 2\,v_3+\tfrac{3}{2}\left(v_1-
u_2+\tfrac{1}{2}\,u_1^2 \right)^2-\tfrac 34\, v_1^2;
\\
\label{9-4} u_t =\left(u_2-\tfrac{3}{2}\,v_1-\tfrac{3}{2}\,u v
-\tfrac{1}{2}\,u^3\right)_x,\qquad v_t=\left(-\tfrac 1
2\,v_2+\tfrac{3}{2}\left(v- u_1+\tfrac{1}{2}\,u^2 \right)^2-\tfrac
34\, v^2\right)_x;
\\
u_t =u_3-3gv_2-3u_1(u_1+ v_1) -\tfrac32\, v_1^2-6v_1g^2 -c_1g^3- 3g^4,   \nonumber\\
v_t =-\tfrac 1 2 v_3-\tfrac34\,c_1u_2+3u_1^2-\tfrac{3}{2}\, v_1^2
- 6u_1g^2+c_1g^3+3g^4,\qquad g=u+v;\label{9-8}
\\
\label{11,1} u_t =u_3-3u_1v_1+(u-3v^2)u_1,\qquad v_t=-\tfrac 1 2
v_3+\tfrac12{u_2}-u_1v-(u-3v^2)v_1;
\\
u_t=u_3-3u_1v_2+uu_1-3u_1v_1^2,  \qquad v_t=-\tfrac 1 2
v_3+\tfrac12u_1-u\,v_1+v_1^3; \label{11,2}
\\
u_t =u_3+\left(k+\sqrt{u^2+v_1}\,\right)\,u_1 , \nonumber\\
v_t =-\tfrac 1 2\, v_3 +\tfrac{3}{8}\frac{(2u\,u_1+ v_2)^2}{u^2 +
v_1}- 3u\,u_2- k(2u^2+v_1)-\tfrac{2}{3}(u^2
+v_1)^{3/2};\label{9-2}
\\
u_t =u_3-\tfrac{3}{4}\,\frac{(2\,v\,v_1+ u_2)^2}{v^2 +u_1}+3v\,v_2 +\tfrac{3}{2}v_1^2 +\tfrac{2}{3}v^3- k(2 v^2+u_1),\nonumber\\
v_t =-\tfrac 1 2 v_3 +\tfrac12 u_2+k\,v_1 ;\label{9-5}
\\
u_t =u_3+\tfrac {u_1(u_1+v_2)}{\sqrt{u+v_1}}-\tfrac{4}{3} u_1 v_1+c_1u_1\sqrt{u+v_1},\nonumber\\
v_t =-\tfrac 1 2 v_3-\tfrac{3}{2} u_2+\tfrac{3}{8}
\frac{(u_1+v_2)^2}{ u +v_1} +\tfrac 23 v_1^2 -\tfrac43
u^2-2u_1\sqrt{u+v_1} -\tfrac23c_1( u +v_1)^{3/2};\label{9-7}
\\
u_t =u_3+u_1\sqrt{\rule{0mm}{3mm}u+ v_1}-k\,u_1,\nonumber\\
v_t =-\tfrac 1 2 v_3- \tfrac{3}{2}u_2+
\tfrac{3}{8}\frac{(u_1+v_2)^2}{u+ v_1}-\tfrac{2}{3}(u+
v_1)^{3/2}+2ku+kv_1; \label{5-1-1-1}
\\ \label{s11a}
u_t=u_3+uv_1+(u^2+v)u_1,\qquad v_t=-\tfrac12
v_3+3u_1u_2-(u^2+v)v_1;
\\
 u_t=u_3+3(u+k)v_2+3u_1(v_1+u^2),\nonumber\\
 v_t=-\tfrac12 v_3-\tfrac32uu_2-\tfrac32(v_1+u^2)^2-\tfrac34 u_1^2+ku^3+\tfrac34 u^4;\label{s11b}
\\
u_t =u_3- \tfrac32\,v_2-\tfrac32\,u_1v_1-\tfrac12u_1^3-3\,u_1(c_1e^u+2c_2e^{2u}),\nonumber\\
v_tv=-\tfrac 1 2 v_3+\tfrac32\left(\tfrac12u_1^2-u_2+v_1+c_1e^u+2c_2e^{2u}\right)^2\nonumber\\
\phantom{v_tv=}{}
-\tfrac34\,v_1^2-\tfrac32\,c_1u_2e^u+\tfrac34\,c_1^2e^{2u}+2c_1c_2e^{3u},\qquad
c_1\ne0\quad \text{or}\quad c_2\ne0;\label{9-9}
\\
u_t =u_3-\tfrac32\,u_1v_2-\tfrac34\,u_1v_1^2+u\,u_1-c^2u_1e^{-2v}, \nonumber\\
v_t =-\tfrac 1 2
v_3+\tfrac14\,v_1^3+{u_1}-u\,v_1+c^2v_1\,e^{-2v};\label{11,1-0}
\\
u_t=u_3-\tfrac{3}{2}u_1v_2-\tfrac{3}{4}u_1v_1^2+u_1\,e^{v}(u_1+2u\,v_1)-\tfrac{1}{3} \,u^2u_1e^{2v}-\tfrac32c_1^2 u_1 \,e^{-2v}, \nonumber\\
v_t =-\tfrac 1 2 v_3+\tfrac{v_1^3}{4}+
u_2\,e^{v}+\tfrac{1}{3}u\, e^{2v}(2u_1+u\,v_1)+\tfrac32c_1^2 v_1
\,e^{-2v},\qquad c_1\ne0;\label{11,1-0-0}
\\
u_t=u_3+3u_2v_1+\tfrac32 u_1v_2+\tfrac94 u_1v_1^2-uu_1e^{2v} -e^{-3v},\nonumber\\
v_t=-\tfrac12 v_3+\tfrac14 v_1^3+(u_1+u v_1)e^{2v};\label{ds2}
\\
 u_t=u_3+3u_2v_1+\tfrac32 u_1v_2+\tfrac94 u_1v_1^2-uu_1e^{2v} -\tfrac14u_1e^{-2v},\nonumber\\
 v_t=-\tfrac12 v_3+\tfrac14 v_1^3+(u_1+u v_1)e^{2v}+\tfrac14v_1e^{-2v};\label{ds2d}
\\
u_t =u_3+\tfrac 3 2 u_1v_2+3u_2v_1+\tfrac 94u_1 v_1^2-c_1^2\,u_1e^{-2v}-\tfrac{1}{2}\,u_1e^{2v}(u^2+c_2), \nonumber\\
v_t =-\tfrac 1 2 v_3+\tfrac 1 4
v_1^3+c_1^2v_1e^{-2v}+\tfrac{1}{2}\,e^{2v}(2uu_1
+u^2v_1+c_2v_1);\label{11,1-1}
\\
u_t=u_3+\tfrac 3 2 u_1v_2+3u_2v_1+\tfrac 94u_1 v_1^2-\tfrac13\,e^{2v}\left(u_1(6u^2+c_1)+4uv_1(2u^2+c_1)\right)\nonumber\\
\phantom{u_t=}{} + e^{v}\left(v_2(2u^2+c_1)+(u_1+2u\,v_1)^2+2c_1v_1^2 \right), \nonumber\\
v_t =-\tfrac 1 2 v_3+\tfrac14{v_1^3}+\tfrac 13\,e^{2v}\left(4
u\,u_1+ (6u^2+c_1)v_1 \right)+\,e^{v}\left(u_2 +
2u_1v_1\right);\label{11,1-2}
\\
u_t =u_3 +\tfrac{3}{2}\,u_1v_2+3u_2v_1+\tfrac{9}{4}\,u_1v_1^2+3uv_2(c_1ue^v+c_2)+c_1(c_1^2-1)u^4e^{3v}\nonumber\\
\phantom{u_t =}{} -\tfrac34\,u^2e^{2v}\big(u_1(1+5c_1^2)+8c_1^2uv_1+2c_2u(1-3c_1^2)\big)- 3c_2^2(u_1+2uv_1) \nonumber\\
\phantom{u_t =}{} +\tfrac32c_1e^v(u_1+2uv_1-2c_2u)^2+\tfrac32c_2v_1(2u_1+3uv_1), \nonumber\\
v_t=-\tfrac 1 2 v_3+\tfrac14 {v_1^3}+\tfrac32c_1e^v(u_2+2u_1v_1)+c_1(1-c_1^2)u^3e^{3v}+6c_1c_2ue^v(v_1-c_2) \nonumber\\
\phantom{v_t=}{} + \tfrac34
ue^{2v}\big(2u_1(1+c_1^2)+uv_1(1+5c_1^2)+2c_2u(1-3c_1^2)\big)+\tfrac32c_2v_1(2c_2-v_1);\label{e1}
\\
u_t =u_3+\tfrac32\,u_1v_2+3u_2v_1+\tfrac94\,u_1v_1^2+3e^v(u^2+c)(v_2+2v_1^2)+\tfrac32\,e^vu_1(u_1+4uv_1)\nonumber\\
\phantom{u_t =}{} -\tfrac32e^{2v}\left((3u^2+c)u_1+4(u^2+c)uv_1\right) ,\nonumber\\
v_t=-\tfrac12
v_3+\tfrac14\,v_1^3+\tfrac32\,e^v(u_2+2u_1v_1)+\tfrac32e^{2v}\left(2uu_1+(3u^2+c)v_1\right);\label{1d-5}
\\
u_t =u_3-\tfrac3 2u_1 v_2 -\tfrac 3 4 u_1 v_1^2 +\tfrac 1 4 u_1^3 -c_1\,e^{-2v}u_1-c_2\,(u_1+2v_1)\,e^{2(u+v)}\nonumber\\
\phantom{u_t =}{} +c_3\, (u_1-2v_1)\,e^{2(v-u)},\nonumber\\
v_t =-\tfrac 1 2 v_3+\tfrac 3 2 u_1u_2-\tfrac 3 4 u_1^2v_1+\tfrac
1 4 v_1^3 + c_1\,e^{-2v}v_1+
\big(c_2\,e^{2(u+v)}-c_3\,e^{2(v-u)}\big) v_1;\label{2-1}
\\
u_t =u_3-\tfrac3 2u_1 v_2 -\tfrac 3 4 u_1 v_1^2 +\tfrac 1 4 u_1^3 +\left(c_2\,e^{u}+c_3\,e^{-u}-3c_1^2\,e^{-2v}\right) u_1,\nonumber\\
v_t =-\tfrac 1 2 v_3+\tfrac 3 2 u_1u_2-\tfrac 3 4 u_1^2v_1+\tfrac 1 4 v_1^3+\left( c_2\,e^{u}-c_3\,e^{-u} \right)u_1\nonumber\\
\phantom{v_t =}{} +\left(3 c_1^2\,e^{-2v}-c_2\,e^{u}-c_3\,e^{-u}
\right)v_1;\label{2-2}
\\
u_t =u_3-\tfrac3 2u_1 v_2 -\tfrac 3 4 u_1 v_1^2 +\tfrac 1 4 u_1^3 +3\,k(u_1^2-2v_2)\,e^{-v}-3(c^2-3k^2)\,u_1\,e^{-2v}\nonumber\\
\phantom{u_t =}{} +8 \,k(k^2-c^2)\,e^{-3v} , \nonumber\\
v_t =-\tfrac 1 2 v_3+\tfrac 3 2 u_1u_2-\tfrac 3 4 u_1^2v_1+\tfrac
1 4
v_1^3+3\,k(u_2-2u_1v_1)\,e^{-v}+3(c^2-3k^2)v_1\,e^{-2v};\label{2-3}
\\
u_t =u_3-\tfrac3 2u_1 v_2 -\tfrac 3 4 u_1 v_1^2 +\tfrac 1 4 u_1^3- 3 c_1^2u_1e^{2(u+v)}+ 3 c_1u_1(u_1+2v_1)e^{u+v} \nonumber\\
\phantom{u_t =}{} +(c_2\,e^{-u}-3c_3^2\,e^{-2v})u_1,  \nonumber\\
v_t =-\tfrac 1 2 v_3+\tfrac 3 2 u_1u_2-\tfrac 3 4 u_1^2v_1+\tfrac 1 4 v_1^3+ 3c_1^2(2u_1+v_1)\,e^{2(u+v)} +3c_1(u_2+u_1^2)e^{u+v} \nonumber\\
\phantom{u_t =}{} -c_2\,
(u_1+v_1)e^{-u}+3c_3^2v_1e^{-2v};\label{2-4}
\\
u_t =u_3-\tfrac3 2u_1 v_2 -\tfrac 3 4 u_1 v_1^2 +\tfrac 1 4 u_1^3+3 c_2\, u_1(u_1+2v_1)\,e^{u+v}-4c_1c_2\,e^{3(u+v)}\nonumber\\
\phantom{u_t =}{} +3 [(c_1-c_2^2)u_1+2c_1v_1]\,e^{2(u+v)},  \nonumber\\
v_t =-\tfrac 1 2 v_3+\tfrac 3 2 u_1u_2-\tfrac 3 4 u_1^2v_1+\tfrac 1 4 v_1^3+3c_2\,(u_2+u_1^2)\,e^{u+v}+4c_1c_2\,e^{3(u+v)}\nonumber\\
\phantom{v_t =}{} +3[2c_2^2u_1-(c_1-c_2^2)v_1]\,e^{2(u+v)};
\label{2-5}
\\
u_t =u_3-\tfrac3 2u_1 v_2 -\tfrac 3 4 u_1 v_1^2 +\tfrac 1 4 u_1^3+\tfrac23\,c_1^2u_1\,e^{-2v} +c_1(2v_2-u_1^2)\,e^{-v}\nonumber\\
\phantom{u_t =}{} -2c_1c_2(u_1+2v_1)\,e^u+3c_2\,u_1(u_1+2v_1)\,e^{u+v}- 3 c_2^2\,u_1\,e^{2(u+v)},\nonumber\\
v_t =-\tfrac 1 2 v_3+\tfrac 3 2 u_1u_2-\tfrac 3 4 u_1^2v_1+\tfrac 1 4 v_1^3-\tfrac23\,c_1^2v_1\,e^{-2v} +c_1(2u_1v_1-u_2)\,e^{-v}\nonumber\\
\phantom{v_t =}{}
-2c_1c_2(u_1-v_1)\,e^u+3c_2\,(u_2+u_1^2)\,e^{u+v}+3
c_2^2\,(2u_1+v_1)\,e^{2(u+v)};\label{2-6}
\\
u_t =u_3-\tfrac3 2u_1 v_2 -\tfrac 3 4\, u_1 v_1^2 +\tfrac 1 4\, u_1^3-3\, u_1\big[c_1^2\,e^{2(u+v)}+c_2^2\, e^{2(v-u)}+2c_1c_2\,e^{2v}\big] \nonumber\\
\phantom{u_t =}{} -3c_3^2\,u_1\,e^{-2v}+3\,c_1u_1(u_1+2v_1)\,e^{u+v}-3\,c_2u_1(u_1-2v_1)\,e^{v-u},  \nonumber\\
v_t =-\tfrac 1 2 v_3+\tfrac 3 2 u_1u_2-\tfrac 3 4 u_1^2v_1+\tfrac 1 4 v_1^3+3c_1^2\,(2u_1+v_1)\,e^{2(u+v)}+6c_1c_2\,v_1\,e^{2v}  \nonumber\\
\phantom{v_t =}{} +3c_2^2
\,(v_1-2u_1)\,e^{2(v-u)}+3\,c_1(u_2+u_1^2)\,e^{u+v}+3c_2\,(u_1^2-u_2)\,e^{v-u}+3c_3^2\,v_1\,e^{-2v};\label{2-7}
\\
u_t =u_3-\tfrac3 2u_1 v_2 -\tfrac 3 4 u_1 v_1^2 +\tfrac 1 4 u_1^3-6c_1^3 \,e^{3(u+v)}-\tfrac34\,c_1^2(5u_1-8v_1)\,e^{2(u+v)}\nonumber\\
\phantom{u_t =}{}
+\tfrac{9}{2}\,c_1u_1(u_1+2v_1)\,e^{u+v}+2c_1^2c_2\,e^{2u+v}+\tfrac23\,c_1c_2^2\,e^{u-v}-
\tfrac12\,c_1c_2(7u_1+12v_1)\,e^{u}\nonumber\\
\phantom{u_t =}{}+c_2(2v_2-u_1^2)\,e^{-v}+\tfrac{11}{12}\,c_2^2\, u_1\,e^{-2v} -\tfrac29\,c_2^3\,e^{-3v} ,  \nonumber\\ v_t =-\tfrac 1 2 v_3+\tfrac 3 2 u_1u_2-\tfrac 3 4 u_1^2v_1+\tfrac 1 4 v_1^3+6c_1^3 \,e^{3(u+v)}+\tfrac34\,c_1^2(18u_1+5v_1)\,e^{2(u+v)}\nonumber\\
\phantom{v_t =}{} +\tfrac{9}{2}\,c_1(u_2+u_1^2)\,e^{u+v}-4c_1^2c_2\,e^{2u+v}+\tfrac23\,c_1c_2^2\,e^{u-v}- \tfrac72\,c_1c_2(u_1-v_1)\,e^{u}\nonumber\\
\phantom{v_t =}{}
+c_2(2u_1v_1-u_2)\,e^{-v}-\tfrac{11}{12}\,c_2^2\,
v_1\,e^{-2v};\label{2-8}
\\
u_t =u_3-\tfrac3 2u_1 v_2 -\tfrac 3 4 u_1 v_1^2 +\tfrac 1 4 u_1^3+c_3e^{-v}(u_1^2-2v_2)+\tfrac 23c_3^2u_1e^{-2v}\nonumber\\
\phantom{u_t =}{} +c_1\left(3u_1e^{u+v}+2c_3e^u \right)(u_1+2v_1)-c_2(3u_1e^{v-u}+2c_3e^{-u})(u_1-2v_1)\nonumber\\
\phantom{u_t =}{}-3u_1(c_1e^{u}+c_2e^{-u})^2e^{2v},\nonumber\\
v_t =-\tfrac 1 2 v_3+\tfrac 3 2 u_1u_2-\tfrac 3 4 u_1^2v_1+\tfrac 1 4 v_1^3+c_3e^{-v}(u_2-2u_1v_1)-\tfrac 23c_3^2v_1e^{-2v}\nonumber\\
\phantom{v_t =}{}  +3c_1^2e^{2(u+v)}(2u_1+v_1)-3c_2^2e^{2(v-u)}(2u_1-v_1)+3c_1e^{u+v}(u_2+u_1^2)\nonumber\\
\phantom{v_t
=}{}-3c_2e^{v-u}(u_2-u_1^2)+2c_2c_3e^{-u}(u_1+v_1)+2c_1c_3e^u(u_1-v_1)+6c_1c_2v_1e^{2v};\label{2-9}
\\
u_t=u_3-\tfrac34\frac{(2g^3-u_2+2gv_1)^2}{u_1-g^2}+3g(u_2-v_2)-6u_1^2-9u_1v_1- \tfrac32v_1^2\nonumber\\
\phantom{u_t=}{} - 3(5g^2+4cg+c^2)u_1-6g^2v_1+2cg^2(8g+3c)+9g^4,\nonumber\\
v_t =-\tfrac12 v_3+\tfrac34\frac{(2g^3-u_2+2gv_1)^2}{u_1-g^2}-3(3g+c)u_2- \tfrac32v_1^2+3(9g^2+8cg+2c^2)u_1\nonumber\\
\phantom{v_t =}{}  +3(6g^2+4cg+c^2)v_1-2cg^2(8g+3c)-9g^4,\qquad
g=u+v;\label{1d-6}
\end{gather}

\begin{remark}
 Systems (\ref{9}),  (\ref{8}), (\ref{5})  and (\ref{s11a}) were proposed in \cite{DS1}, where system (\ref{s11a})
is given with a misprint. System (\ref{1d}) was presented in
\cite{D-S}.
\end{remark}

\begin{remark}
Ten pairs of integrability conditions (for $\rho_0$--$\rho_9$ and
$\t\rho_0$--$\t\rho_9$) have been verif\/ied for each system
(\ref{9})--(\ref{1d-6}), and
 nontrivial higher conserved densities with orders 2, 3, 4 and 5 have been found.
 \end{remark}

\begin{remark}
 It is shown in \cite{FPM} that systems (\ref{1d}) and (\ref{9-4}) are unique divergent systems of the form (\ref{sy2a})
that satisfy the integrability conditions.
\end{remark}

\begin{remark}
System (\ref{9-9}) is a modif\/ication of (\ref{9-40}), systems
(\ref{2-1})--(\ref{2-9}) are modif\/ications of (\ref{1}).
\end{remark}

\begin{remark}
Canonical densities for system (\ref{ds2}) depend on the nonlocal
variable $w=D_x^{-1}e^{-v}$.
\end{remark}

\begin{remark}
Many of the systems possess discrete symmetries. They are:

$u\to -u$ for (\ref{9}), (\ref{9-0}), (\ref{1}), (\ref{1d}),
(\ref{9-2}), (\ref{s11a}) and (\ref{11,1-1});

$u\to -u$, $v\to v+\pi i$ for (\ref{11,1-0-0});

$u\to iu$, $v\to v-\frac i2 \pi$, $c_1\to -c_1$ for
(\ref{11,1-2});

$\{u\to -u,\,v\to v+\pi i\}\cup\{v\to v+\pi i,\,
c_1\to-c_1\}\cup\{u\to-u,\, c_1\to-c_1\}$ for (\ref{e1});

$\{u\to -u,\, v\to v+\pi i\}\cup\{u\to iu,\, v\to v -\frac i2
\pi\}$ for (\ref{1d-5});

$u\to -u$, $c_2\to c_3$, $c_3\to c_1$ for (\ref{2-2});

$u\to -u$, $k\to-k$ for (\ref{2-3});

$u\to -u$, $c_1\to c_2$, $c_2\to c_1$ for (\ref{2-7});

$u\to -u$, $c_3\to-c_3$, $c_2\to c_1$, $c_1\to c_2$ for
(\ref{2-9}).

Also, systems  (\ref{9}), (\ref{9-0}), (\ref{1}), (\ref{1d}) and
(\ref{11,1-1})  preserve the real shape under the transformation
$u\to iu$. System (\ref{2-3})  keeps the real shape under the
transformation $u\to iu$, $k\to ik$.
\end{remark}

\subsection{Example of computations}\label{sec3.1}

Let us consider the simplest case of system (\ref{sys0}):
\begin{gather}\label{ex1}
u_t=u_3+f_1(u,v)u_1+f_2(u,v)v_1,\qquad v_t=a
v_3+g_1(u,v)u_1+g_2(u,v)v_1,
\end{gather}
where $a(a-1)\ne0$. Formulas (\ref{dens}) are reduced now to the
following
\begin{gather}
 \rho_0=0,\qquad \t\rho_0=0,\qquad \rho_1=-\frac{1}{3}\,f_1,\qquad \t\rho_1=-\frac{1}{3a}\,g_2.   \label{dens1}
\end{gather}
The further canonical densities read
\begin{gather}
\rho_2=-\frac13(f_{1,u}u_1+f_{2,u}v_1)+\frac13D_xf_1,\qquad \t\rho_2=-\frac{1}{3a}(g_{1,v}u_1+g_{2,v}v_1)+\frac{1}{3a}D_xg_2,\nonumber\\
\tau_2=\frac13(f_{1,u}u_1+f_{2,u}v_1),\qquad
\t\tau_2=\frac{1}{3a}(g_{1,v}u_1+g_{2,v}v_1),\label{dens2}
\end{gather}
where indices after commas denote derivatives.

The f\/irst integrability condition (\ref{con1}) for $\rho_1$ can
be split with respect to $u_3$, $v_3$, $u_2$ and $v_2$. This
provides the following equations
\begin{gather*}
f_{1,uv}=0,\!\!\qquad f_{1,uuu}=0,\!\!\qquad
f_{1,vvv}=0,\!\!\qquad \text{or}\!\!\qquad
f_1(u,v)=c_1u^2+c_2u+c_3v^2+c_4v+c_5.
\end{gather*}
Analogously, condition (\ref{con1}) for $\t\rho_1$ implies
$g_2(u,v)=b_1u^2+b_2u+b_3v^2+b_4v+b_5$. It is obvious from
(\ref{dens2}) that the second integrability conditions
(\ref{law2}) are $\tau_2\in\I D_x$ and $\t\tau_2\in\I D_x$. These
conditions provide
 $f_{2,uu}=g_{1,vv}=0$ or $f_2(u,v)=uf_3(v)+f_4(v)$, $g_1(u,v)=vg_3(u)+g_4(u)$.

Thus, system (\ref{ex1}) takes the following form:
\begin{gather}
 u_t=u_3+(c_1u^2+c_2u+c_3v^2+c_4v+c_5)u_1+(uf_3(v)+f_4(v))v_1,\nonumber\\
 v_t=a v_3+(vg_3(u)+g_4(u))u_1+(b_1u^2+b_2u+b_3v^2+b_4v+b_5)v_1.\label{ex2}
\end{gather}

Now one can obtain $\theta_2=D_x^{-1}D_t\rho_2$ and
$\t\theta_2=D_x^{-1}D_t\t\rho_2$ in an explicit form. The
expressions~$D_t\rho_1$ and $D_t\t\rho_1$ are not the total
derivatives yet:
\begin{gather*}
D_t\rho_1=D_xh_1(u_i,v_j)+R_1(u,v,u_1,v_1)=D_x\theta_1,\\
D_t\t\rho_1=D_x\t h_1(u_i,v_j)+\t R_1(u,v,u_1,v_1)=D_x\t\theta_1.
\end{gather*}
Therefore, we have set $\theta_1=h_1(u_i,v_j)+q_1(u,v)$,
$\t\theta_1=\t h_1(u_i,v_j)+\t q_1(u,v)$, where $q_1$ and $\t q_1$
are unknown functions and $D_xq_1=R_1$, $D_x\t q_1=\t R_1$. This
trick allows us to evaluate $\rho_4$, $\tau_4$, $\t\rho_4$,
$\t\tau_4$ and verify the fourth integrability conditions
(\ref{law2}). These conditions imply $f_3''=g_3''=0$, hence
\[
f_3=a_1v+a_2,\qquad  g_3=a_3u+a_4.
\]
To simplify the further analysis one must list all irreducible
cases of $f_1$ (or $g_2$). Let us take
$f_1=c_1u^2+c_2u+c_3v^2+c_4v+c_5$ for def\/initeness.

\begin{lemma}\label{lemma4} Using complex dilatations of $u$ and $v$, translations $u\to u+\lambda_1$, $v\to v+\lambda_2$ and the Galilei transformation
$u_t\to u_t+\alpha u_x$, $v_t\to v_t+\alpha v_x$ one can reduce
$f_1$ to one of the following forms:
\begin{gather*}
1)\ u^2+v^2;\quad 2)\ u^2+\alpha v;\quad 3)\ v^2+\alpha u;\quad
4)\ u+v;\quad 5)\ u;\quad 6)\ v;\quad 7)\ f_1=0,
\end{gather*}
where $\alpha$ is any constant. Moreover, in the cases {\rm 4--7}
the function $g_2$ must be linear $(b_1=b_3=0)$ because otherwise
the permutation $u\leftrightarrow v$ gives one of the cases {\rm
1--3}.
\end{lemma}

In cases 1 and 3 contradictions follow from the integrability
conditions (\ref{con1}) with $n=1,3$ and (\ref{law2}) with
$n=2,4$. In case 2 the integrability conditions (\ref{con1}) with
$n=1,3,5$ and (\ref{law2}) with  $n=2,4$ are satisf\/ied if\/f
system (\ref{ex2}) is reduced to a pair of independent equations.
Thus, a nontrivial integrable system (\ref{ex1}) must belong to
the following class:
\begin{gather}
 u_t=u_3+(c_2u+c_4v)u_1+(u(a_1v+a_2)+f_4(v))v_1,\nonumber\\
 v_t=a v_3+(v(a_3u+a_4)+g_4(u))u_1+(b_2u+b_4v+b_5)v_1,\label{ex3}
\end{gather}
and only the following cases are possible:
\[
4)\ c_2=c_4=1;\quad 5)\ c_2=1,\ c_4=0;\quad 6)\ c_2=0,\
c_4=1;\quad 7)\ c_2=c_4=0.
\]
In case 4 the integrability conditions (\ref{con1}) with
$n=1,\dots,5$ provide the functions $g_4=k_1u+k_2$,
$f_4=k_3v+k_4$, the coef\/f\/icients $a_3=0$, $a_4 = -1+a_2+b_2$,
$b_4 = b_2=(a+1)a_2-2a-1$ and the following equations:
\begin{gather*}
  (a+1)(2a_2+aa_2-2a-1)=0,\!\!\qquad a^2(a_2-2)-a(4a_2^2-5-7a_2)+16a_2-14a_2^2-3=0,\\
 a^5(a_2-2)-a^4(4a_2^2-7a_2-6)-2a^3(16a_2^2-5a_2-37)\\
\qquad{}-a^2(177a_2^2-224a_2-41)-a(236a_2^2-353a_2+102)-a_2(37a_2-53)
-17=0.
\end{gather*}
Using the package Groebner in Maple, one can obtain  $a_2 =(1
-a)/3$, $a^2+7a+1=0$ or $a=(3c-7)/2$, $c^2=5$. Then, the remaining
coef\/f\/icients are also determined and we obtain
\begin{gather}
 u_t= u_3+(u+v)u_1+ \tfrac12\big((3-c)u+(5c-11)v\big)v_1, \qquad c^2=5,\nonumber\\
 v_t= \tfrac12v_3(3c-7)+\big((c+2)u-v\big)u_1+\tfrac12(c-3)(u+v)v_1.\label{ex4}
\end{gather}
The following substitution
\[
V = \tfrac16(c+1)(v-u),\qquad  U =  \tfrac{1}{12}(c+3)u+ \tfrac16v
\]
reduces system (\ref{ex4}) to  the third Drinfeld--Sokolov system
\setcounter{equation}{45}
\begin{subequations}
\begin{gather}
U_t= -8U_3+ 3V_3+6(V-8U)U_1+12UV_1, \nonumber\\
 V_t= 12U_3-2V_3+48VU_1+12(2U-V)V_1\label{ex5}
\end{gather}
that has been presented f\/irst in \cite{DS1}. Scaling
\[
t\to -\tfrac12 t,\qquad  U =\to \tfrac16 U,\qquad  V =\to
-\tfrac13V
\]
gives more symmetric form of system (\ref{ex5})
\begin{gather}
 U_t= 4U_3+ 3V_3+(4U+V)U_1+2UV_1, \nonumber\\
 V_t= 3U_3+V_3-4VU_1-2(V+U)V_1\label{ex6}
\end{gather}
\end{subequations}
that was found in~\cite{Fo}.

In case 5 the equations $a_1=a_2=0$, $g_4'f_4'=0$, $g_4'''=0$
follow from  the integrability conditions~(\ref{con1}) with
$n=1,\dots,5$. This implies $f_4\ne0$ because otherwise the
f\/irst equation of (\ref{ex3}) will be independent. Hence, there
are two branches (1) $f_4=1$ and (2) $f_4'\ne0$, $g_4'=0$. Along
the f\/irst branch, if one use additionally the integrability
conditions (\ref{con1}) with $n=7$ and solves a large polynomial
system for constants, one can obtain the following system:
\begin{gather*}
u_t=u_3+uu_1+v_1, \qquad v_t=-2v_3-uv_1,
\end{gather*}
that can be transformed to (\ref{s2}) by a scaling.

Along the second branch the integrability conditions (\ref{con1})
with $n=1,\dots,5$ provide  the following system
\begin{gather*}
u_t=u_3+uu_1- vv_1, \qquad v_t=-2v_3-uv_1,
\end{gather*}
that is equivalent to (\ref{s1}).

Case 6 is symmetric to case 5: one can obtain $f_4'''=0$,
$g_4''=0$, $a_3=a_4=b_2=0$ from   the integrability conditions
(\ref{con1}) with $n=1,\dots,5$. Hence $g_4\ne0$ and we have two
branches $g_4=1$ or $g_4=u$. Using the additional integrability
conditions (\ref{con1}) with $n=6,7$ one can obtain
equations~(\ref{s1}) and~(\ref{s2}).

There are many branches in case 7 but all of them provide linear
or triangular systems only.

As one can see, classif\/ication of integrable systems of the form
(\ref{ex1}) is a suf\/f\/iciently laborious task.
System~(\ref{sy2a}) contains eight unknown functions depending on
four variables, therefore classif\/ication of these systems is
much  more dif\/f\/icult.

\section[Differential substitutions]{Dif\/ferential substitutions}\label{sec4}

A dif\/ferential substitution  is a pair of  equations
\begin{gather}\label{difs}
u=f(U,V,U_x,V_x,\dots,U_n,V_n),\qquad
v=g(U,V,U_x,V_x,\dots,U_n,V_n),
\end{gather}
where $f$ and $g$ are some smooth functions.

\begin{definition} If for any solution $(U,V)$ of a system $(\Sigma)$ formulas (\ref{difs}) provide a solution
$(u,v)$ of system (\ref{sys0}), then one says that system
(\ref{sys0})  admits substitution (\ref{difs}).
\end{definition}

In all cases that we know, the new systems $(\Sigma)$ belong to
the same class (\ref{sys0})
\begin{gather}\label{sys2}
U_t=U_{xxx}+P(U,V,U_x,V_x,U_{xx},V_{xx}),\qquad
V_t=a\,V_{xxx}+Q(U,V,U_x,V_x,U_{xx},V_{xx}), \tag{$S$}
\end{gather}
with some  smooth functions $P$ and $Q$. There exist some
group-theoretical explanation of this fact for KdV type equations
\cite{DS2}. Our attempts to introduce another parameter $a'\ne a$
in (\ref{sys2}) had no success.

Substituting (\ref{difs}) into (\ref{sys0}) one obtains the
following equations
\begin{gather}
 \big(D_x^3f+F(f,g,D_xf,D_xg,D_x^2f,D_x^2g)-\p_tf\big)\cond{S}=0, \nonumber\\
 \big(aD_x^3g+G(f,g,D_xf,D_xg,D_x^2f,D_x^2g)-\p_t g\big)\cond{S}=0.\label{deff}
\end{gather}
It is obvious that transition to the manifold ($S$) in
(\ref{deff}) is equivalent to a replacement of $\p_t$ by  the
evolutionary dif\/ferentiation $D_t$ performed in accordance with
(\ref{sys2}):
\begin{gather}
D_tf=D_x^3f+F(f,g,D_xf,D_xg,D_x^2f,D_x^2g),\nonumber\\
D_tg=aD_x^3g+G(f,g,D_xf,D_xg,D_x^2f,D_x^2g),\label{def}
\end{gather}
where
\begin{gather*}
D_tf=\sum_{i=1}^{n} \frac{\p f}{\p U_i}D_x^i(U_3+P)+\sum_{i=1}^{n}
\frac{\p f}{\p V_i}D_x^i(aV_3+Q).\nonumber
\end{gather*}

Another way to obtain (\ref{def}) is to dif\/ferentiate equations
(\ref{difs}) with respect to $t$ in accordance with (\ref{sys0})
and (\ref{sys2}) and exclude $u$ and $v$ by using (\ref{difs}).
This algorithm and many others are coded in Maple, see for example
\cite{JetM}.

To f\/ind the admissible functions $f$, $g$, $P$, $Q$  from
(\ref{def})  one can use the following easily provable formula:
\begin{gather*}
\frac{\p}{\p U_k}D_x^mf=\sum_{s=0}^{m}
\binom{m}{s}D_x^{m-s}\frac{\p f}{\p U_{k-s}},\qquad \frac{\p f}{\p
U_{-i}}\equiv 0\qquad \text{for} \quad i>0,
\end{gather*}
and the analogous formula for $\p/\p V_k$. Dif\/ferentiating (\ref{def})
with respect to $U_{n+3}$ and $V_{n+3}$, one obtains
\begin{gather}\label{x1}
\frac{\p f}{\p V_n}=0,\qquad\frac{\p g}{\p U_n}=0.
\end{gather}
Other corollaries of (\ref{def}) are too cumbersome to consider
them in the general form.

Let us consider, as an example, the f\/irst order dif\/ferential
substitutions for system (\ref{s1})
\begin{gather}\label{DS}
u_t =u_3+v\, u_1,\qquad v_t=-\tfrac 1 2 v_3+u\,u_1-v\,v_1.
\end{gather}
According to (\ref{x1}) one has $f=f(U,V,U_1),\,g=g(U,V,V_1)$,
hence equations (\ref{def}) now read
\begin{gather}
D_x^3f-f_U(U_3+P)-f_V(Q-V_3/2)-f_{U_1}(U_4+D_xP)+g\,D_x f=0, \nonumber\\
g_U(U_3+P)+g_V(Q-V_3/2)+g_{V_1}(D_xQ-V_4/2)+\tfrac12\,D_x^3g-f\,D_xf+
g\,D_xg=0.\label{x3}
\end{gather}
Dif\/ferentiating  (\ref{x3}) with respect to $U_3$ and $V_3$ one
can obtain four equations:
\begin{gather}
 \frac{\p f}{\p U_1}\frac{\p P}{\p U_2}=3D_x\frac{\p f}{\p U_1},\qquad \frac{\p f}{\p U_1}\frac{\p P}{\p V_2}=\frac{3}{2}\frac{\p f}{\p V}, \nonumber\\
 \frac{\p g}{\p V_1}\frac{\p Q}{\p U_2}=-\frac{3}{2}\frac{\p g}{\p U},\qquad \frac{\p g}{\p V_1}\frac{\p Q}{\p V_2}=-\frac{3}{2}D_x\frac{\p g}{\p V_1}.\label{x2}
\end{gather}

Let us consider some corollaries of these equations.
\begin{enumerate}\itemsep=0pt
\item If $\p f/\p U_1=0$ and $\p g/\p V_1=0$, then $u=f(U)$, $
v=g(V)$ is a trivial point transformation.

\item If $\p f/\p U_1=0$, then $u=f(U)$ and one can set $f(U)=U$
by modulo of the point transformation. In this case $P=gU_1$ from
the f\/irst of equations (\ref{x3}).

\item If $\p g/\p V_1=0$, then $v=g(V)$ and one can set $g(V)=V$
by modulo of the point transformation.  In this case
$Q=fD_xf-VV_1$ from the second of equations (\ref{x3}).

\item If $(\p f/\p U_1)(\p g/\p V_1)\ne0$, then one can f\/ind $P$
and $Q$ as polynomials of $U_2$ and $V_2$ from  equations
(\ref{x2}).
\end{enumerate}

Investigation of cases 2--4 provides seven nontrivial solutions of
equations (\ref{x3}) (see below (\ref{9}) $\to$ (\ref{9-0}),
\dots, (\ref{9}) $\to$ (\ref{9-5})).

Note that  {\it integrable} system (\ref{5})  admits strange
dif\/ferential substitutions that generate {\it non-integrable}
systems. For example, system (\ref{5}) admits the following
dif\/ferential substitution:
\begin{gather}
u=\tfrac32 V_2-\tfrac34 V_1^2-\tfrac32 U_1e^V,\nonumber\\
v=\tfrac94\left(-V_4+V_1V_3+V_1^2V_2-\tfrac14V_1^4-U_1^2e^{2V}+e^V(U_3+2U_2V_1+3U_1V_2)\right),\label{stran}
\end{gather}
so that the functions $U$ and $V$ satisfy the following system:
\begin{gather*}
U_t=U_3+\tfrac32U_2V_1+\tfrac34U_1V_1^2-U_1e^{-V}f''(U)+f(U), \\
V_t=-\tfrac12 V_3+\tfrac14V_1^3+\tfrac32e^V(U_2+U_1V_1)-f'(U)
\end{gather*}
with arbitrary function $f$. This system does not satisfy the
integrability conditions (\ref{law}).  To comprehend this unusual
phenomenon we evaluate $V_2$, $V_3$ and $V_4$ from the f\/irst
equation (\ref{stran})
\begin{gather*}
V_2=\tfrac{2}{3}u+U_1e^V+\tfrac12 V_1^2,\qquad
V_3=\tfrac{2}{3}u_1+D_x(U_1e^V)+ V_1V_2,\\
V_4=\tfrac{2}{3}u_2+D_x^2(U_1e^V)+D_x( V_1V_2),
\end{gather*}
and substitute them into the second one. The result is
\[
v=-u^2-\tfrac32u_2.
\]
It is easily verif\/ied that the obtained constraint is a
reduction of system  (\ref{5}) into the single KdV equation
$u_t=-1/2u_3-uu_1$. This means that using a substitution like
(\ref{stran}) we are trying to construct an integrable system from
the  single KdV equation. There are other such examples for
system (\ref{5}). Note that the reduction obtained above follows
from the reduction $u={\rm const}$ for system (\ref{8}) (see
(\ref{8}) and (\ref{8}) $\to$ (\ref{5})).

To organize the presented list of systems we have computed
admissible dif\/ferential substitutions for each system and
present the results in this section. The formula
\begin{gather}\label{exx}
u'=f(u,v,u_x,v_x,\dots),\qquad v'=g(u,v,u_x,v_x,\dots) \tag*{(A)
$\to$ (B)}
\end{gather}
will denote that if $u'$ and $v'$ are substituted into system (A),
then system (B) follows for $u$ and $v$. We say in this case that
system (B) is obtained from system (A) by the dif\/ferential
substitution.

Substitution \ref{exx} establishes an interrelation between the
sets of solutions of systems~(A) and~(B): $(u,v)\mapsto (u',v')$
is a single valued map. And conversely, if for some
solution~$(u',v')$ of system~(A) one solves the system of two
ordinary dif\/ferential equations~\ref{exx} for~$u$ and~$v$, then
one or more solutions of system~(B) are obtained. Of course,
explicit solutions can be obtained very rarely when the
substitution  is linear or invertible (see below).

Let us consider the following simple example:
\begin{gather*}
u'=u_1,\qquad v'=v_1 .  \tag*{(\ref{1d}) $\to$ (\ref{1})}
\end{gather*}
This substitution is possible for any divergent system
\begin{gather*}
u_t=\big(u_2+F(u,v,u_1,v_1)\big)_x,\qquad v_t=\left(-\tfrac12
v_2+G(u,v,u_1,v_1)\right)_x.
\end{gather*}
It produces the system $u_t=u_3+F(u_1,v_1,u_2,v_2)$,
$v_t=-v_2/2+G(u_1,v_1,u_2,v_2)$ without~$u_0$ and~$v_0$. The
inverse transformation is quasi-local $u=D_x^{-1}u'$,
$v=D_x^{-1}v'$. This is  a well known fact, that is why the
substitutions $(u,v)\to(u_1,v_1)$ are not written for the
divergent systems below.  In some cases analogous substitutions
are not so obvious and we present them likewise (\ref{9}) $\to$
(\ref{9-0}), for example.

\begin{theorem} Differential substitutions presented below connect all systems from the list of Section~{\rm \ref{sec3}} with systems \eqref{s1} and \eqref{s2}.
Systems~\eqref{s1} and \eqref{s2} are also implicitly connected
with each other.
\end{theorem}

The proof can be obtained by a direct verif\/ication.

List of the substitutions:
\begin{gather}
u'=u,\qquad v'=v_1;   \tag*{(\ref{9}) $\to$ (\ref{9-0})}
\\
u'=\tfrac{3}{\sqrt{2}}\,(u_2-u_1v_1),\qquad
v'=\tfrac32v_2-\tfrac34(u_1^2+v_1^2);  \tag*{(\ref{9}) $\to$
(\ref{1})}
\\
u'=\tfrac{3}{\sqrt{2}}\,(u_1-u\,v),\qquad
v'=\tfrac32v_1-\tfrac34(u^2+v^2);  \tag*{(\ref{9}) $\to$
(\ref{1d})}
\\
u'=\tfrac{3}{4}\,\sqrt{2}\left(u_1^2- 2u_2+ 2v_1\right),\qquad
v'=\tfrac{3}{2}\,v_1; \tag*{(\ref{9}) $\to$ (\ref{9-40})}
\\
u'=\tfrac{3}{4}\,\sqrt{2}\left(u^2- 2u_1+ 2v\right),\qquad
v'=\tfrac{3}{2}\,v;   \tag*{(\ref{9}) $\to$ (\ref{9-4})}
\\
u'=3\sqrt{2}\left((u+v)^2-u_1\right)+\tfrac{3}{16}\sqrt{2}\,c_1^2,\qquad
v'=3v_1+\tfrac{3}{2}c_1(u+v);  \tag*{(\ref{9}) $\to$ (\ref{9-8})}
\\
u'=\sqrt{2}\, u,\qquad v'=-3v_1+u-3v^2;  \tag*{(\ref{9}) $\to$
(\ref{11,1})}
\\
u'=u,\qquad v'=k+\sqrt{u^2+v_1};  \tag*{(\ref{9}) $\to$
(\ref{9-2})}
\\
u'= \sqrt{v^2+u_1},\qquad v'=v-k;  \tag*{(\ref{9}) $\to$
(\ref{9-5})}
\\
u'=\sqrt{2}\left(\tfrac43 u+\tfrac{3}{16}c_1^2\right),\qquad
v'=\frac{u_1+v_2}{\sqrt{u+v_1}}-\tfrac43 v_1+c_1\sqrt{u+v_1};
\tag*{ (\ref{9}) $\to$ (\ref{9-7})}
\\
u'=\tfrac{3}{2}\,\sqrt{2}\left(\tfrac12u_1^2-
u_2+v_1+c_1e^u+2c_2e^{2u}\right),\qquad
v'=\tfrac{3}{2}\,(v_1+c_1e^u); \tag*{(\ref{9}) $\to$ (\ref{9-9})}
\\
u'=\sqrt{2}\,u,\qquad
v'=-\tfrac{3}{2}\,v_2-\tfrac34\,v_1^2+u-c^2e^{-2v};
\tag*{(\ref{9}) $\to$ (\ref{11,1-0})}
\\
u'=2c_1u,\qquad v'=-\tfrac32
v_2-\tfrac34v_1^2-\tfrac13u^2e^{2v}-\tfrac32c_1^2\,e^{-2v}+e^{v}(u_1+2uv_1);
\tag*{(\ref{9}) $\to$ (\ref{11,1-0-0})}
\\
u'=\sqrt{3}\,u_1e^{v}+2c_1u,\nonumber\\
 v'=\tfrac 32\, v_2-\tfrac34\,v_1^2+2\sqrt{3}\,c_1v_1e^{-v}- \tfrac12\,e^{2v}(u^2+c_2)-c_1^2e^{-2v};
\tag*{(\ref{9}) $\to$ (\ref{11,1-1})}
\\
u'=\tfrac13\sqrt{2}\,e^{v}(3u_1+c_1e^{v}+ 2u^2e^{v}),\nonumber\\
v'=\tfrac 3 2 v_2-\tfrac 34\,v_1^2-
u_1e^{v}+\tfrac13\,e^{2v}(c_1-2u^2);  \tag*{(\ref{9}) $\to$
(\ref{11,1-2})}
\\
u' =\tfrac32\sqrt{2}e^{v}(u_1+2c_2u+c_1u^2e^{v}),\nonumber\\
v' =\tfrac 32\, v_2-\tfrac
34\,v_1^2-\tfrac32c_1u_1e^{v}+3c_2v_1-\tfrac
34(2c_2+c_1ue^v)^2-\tfrac 34u^2e^{2v};
   \tag*{(\ref{9}) $\to$ (\ref{e1})}
\\
u'=\tfrac32\sqrt{2}e^{v}\big(u_1+(c+u^2)e^{v}\big),\qquad
v'=\tfrac 32 v_2-\tfrac 34\,v_1^2-\tfrac32u_1e^{v}+\tfrac
32(c-u^2)e^{2v};\!\!
 \tag*{(\ref{9}) $\to$ (\ref{1d-5})}
\\
u' =\tfrac32\,\sqrt{2}\,(u_2-u_1v_1)+\sqrt{2}\left(c_2\,e^{u}-c_3\,e^{-u}-3c_1u_1\,e^{-v}\right), \nonumber\\
v' =
\tfrac32\,v_2-\tfrac34\,(u_1^2+v_1^2)+c_2\,e^{u}+c_3\,e^{-u}-6c_1v_1\,e^{-v}-3c_1^2\,e^{-2v};
 \tag*{(\ref{9}) $\to$  (\ref{2-2})}
\\
u' =\tfrac32\,\sqrt{2}\,(u_2-u_1v_1)-3\sqrt{2}(cu_1+2kv_1)\,e^{-v}-6ck\sqrt{2}\,e^{-2v},\nonumber \\
v' =
\tfrac32\,v_2-\tfrac34\,(u_1^2+v_1^2)-3(ku_1+2cv_1)\,e^{-v}-3(c^2+k^2)\,e^{-2v};
 \tag*{(\ref{9})  $\to$  (\ref{2-3})}
\\
u' =\tfrac{3}{\sqrt{2}}\,(u_2-u_1v_1) -\sqrt{2}\left(c_2\,e^{-u}+6c_1c_3\,e^{u}-3c_1u_1\,e^{u+v}+3c_3u_1\,e^{-v}\right),  \tag*{(\ref{9})  $\to$  (\ref{2-4})}\\
v' =
\tfrac32\,v_2-\tfrac34\,(u_1^2+v_1^2)+c_2\,e^{-u}-3c_1^2\,e^{2(u+v)}-3c_1u_1\,e^{u+v}
-6c_3v_1\,e^{-v}-3c_3^2\,e^{-2v};\nonumber
\\
u' =\tfrac{3}{\sqrt{2}}\,(u_2-u_1v_1)+ 3\sqrt{2}\left(c_1\,e^{2(u+v)}+c_2u_1\,e^{u+v}\right), \nonumber\\
v' =
\tfrac32\,v_2-\tfrac34\,(u_1^2+v_1^2)+3(c_1-c_2^2)\,e^{2(u+v)}-3c_2u_1\,e^{u+v};
  \tag*{(\ref{9})  $\to$  (\ref{2-5})}
\\
u' =\sqrt{2}\left(\tfrac32\,(u_2-u_1v_1)-\tfrac23\,kc_1^2\,e^{-2v}+3c_2u_1\,e^{u+v}+2kc_1c_2\,e^u\right)\nonumber\\
\phantom{u' =}{} +c_1\sqrt{2}\,(ku_1+2v_1)e^{-v},\qquad k^2=1, \nonumber\\
v' = \tfrac32\,v_2-\tfrac34\,(u_1^2+v_1^2)-\tfrac23\,c_1^2\,e^{-2v}-3c_2^2\,e^{2(u+v)}-3c_2u_1\,e^{u+v}\nonumber\\
\phantom{v' =}{} +2c_1c_2\,e^u+c_1(u_1+2kv_1)e^{-v};
 \tag*{(\ref{9})  $\to$  (\ref{2-6})}
\\
u' =\tfrac32\,\sqrt{2}\,(u_2-u_1v_1)+3\sqrt{2}\left(c_1\,e^{u+v}+c_2\,e^{v-u}-c_3\,e^{-v}\right)u_1 \nonumber\\
\phantom{u' =}{} +6c_3\sqrt{2}\,(c_2\,e^{-u}-c_1\,e^u),\nonumber\\
v' = \tfrac32\,v_2-\tfrac34\,(u_1^2+v_1^2)-3c_1^2\,e^{2(u+v)}-3c_2^2\,e^{2(v-u)}\nonumber\\
\phantom{v' =}{} +3u_1(c_2\,e^{v-u}-
c_1\,e^{u+v})-6c_3v_1\,e^{-v}-6c_1c_2\,e^{2v}-3c_3^2\,e^{-2v};
 \tag*{(\ref{9})  $\to$  (\ref{2-7})}
\\
u' = \tfrac32\,\sqrt{2}\,(u_2-u_1v_1)+\tfrac13\sqrt{2}\,c_2^2\,e^{-2v}-\tfrac12\sqrt{2}\,c_2(u_1-4v_1)\,e^{-v}\nonumber\\
\phantom{u' =}{}-2\sqrt{2}\,c_1c_2\,e^u+\tfrac32\,\sqrt{2}\,c_1(2c_1\,e^{2(u+v)}+3u_1\,e^{u+v}), \nonumber\\
v' = \tfrac32\,v_2-\tfrac34\,(u_1^2+v_1^2)-\tfrac{15}{4}\,c_1^2\,e^{2(u+v)}-\tfrac92\,c_1u_1\,e^{u+v}+\tfrac52\,c_1c_2\,e^u \nonumber\\
\phantom{v' =}{}
-\tfrac{5}{12}\,c_2^2\,e^{-2v}+c_2(u_1-v_1)e^{-v};\tag*{(\ref{9})
$\to$  (\ref{2-8})}
\\
u' =\tfrac32\,\sqrt{2}\,(u_2-u_1v_1)+3\sqrt{2}\,u_1\left(c_1\,e^{u+v}+c_2\,e^{v-u}\right)-\tfrac23 \sqrt{2}\,c_3^2\,e^{-2v}\nonumber\\
\phantom{u' =}{} +2\sqrt{2}\,c_3(c_2\,e^{-u}-c_1\,e^u)- \sqrt{2}\,c_3\,(u_1+2v_1)e^{-v}, \nonumber\\
v' = \tfrac32\,v_2-\tfrac34\,(u_1^2+v_1^2)-3e^{2v}(c_1\,e^{u}+c_2\,e^{-u})^2-c_3(u_1+2v_1)e^{-v} \nonumber\\
\phantom{v' =}{} +2c_3(c_2\,e^{-u}-c_1\,e^u) -
\tfrac23\,c_3^2\,e^{-2v}+3u_1\big(c_2\,e^{v-u}-c_1\,e^{u+v}\big);
  \tag*{(\ref{9})  $\to$  (\ref{2-9})}
\\
u'=\tfrac{3}{\sqrt{2}}\frac{u_2-3gv_1-2g^3}{\sqrt{u_1-g^2}}- 3\sqrt{2}\,(3g+2c)\sqrt{u_1-g^2},\nonumber\\
 v'=3(v_1-2cg-c^2),\qquad g=u+v;
 \tag*{(\ref{9})  $\to$  (\ref{1d-6})}
\\
u'=u,\qquad v'=v_1; \tag*{(\ref{8}) $\to$ (\ref{8-2})}
\\
u'=\tfrac32 u_2+u^2+v,\qquad v'=u,   \tag*{(\ref{8}) $\to$
(\ref{5})}
\\
\mbox{this substitution is invertible, see below (\ref{5}) $\to$
(\ref{8})};\nonumber
\\
u'=u_1+\tfrac12v^2,\qquad v'=v-k; \tag*{(\ref{8}) $\to$
(\ref{5-1-1})}
\\
u' =4u_2+u_1\left( \tfrac{16}{3}\sqrt{u+v_1}-2c_1 \right)+\tfrac{16}{9}u^2+\tfrac{1}{2}c_1^2u ,\nonumber\\
v' =-
\tfrac{u_1+v_2}{\sqrt{u+v_1}}-\tfrac{4}{3}v_1+c_1\sqrt{u+v_1};
\tag*{(\ref{8}) $\to$ (\ref{9-7})}
\\
u'=u_1,\qquad v'=v;  \tag*{(\ref{8-1}) $\to$ (\ref{8})}
\\
u'=u_1,\qquad v'=v_1;  \tag*{(\ref{8-1}) $\to$ (\ref{8-2})}
\\
u'=\tfrac32u_3+2uu_1+v_1,\qquad v'=u, \tag*{(\ref{8-1}) $\to$
(\ref{5})}
\\
\mbox{this substitution has the quasi-local  inverse substitution,
see below (\ref{5}) $\to$ (\ref{8-1})}; \nonumber
\\
u'=u_2+vv_1,\qquad v'=v-k; \tag*{(\ref{8-1}) $\to$ (\ref{5-1-1})}
\\
u'=3u_3-6(u_1v_1+vu_2)+2uu_1,\qquad v'=3v_1-3v^2+u;
\tag*{(\ref{8-1}) $\to$ (\ref{11,1})}
\\
u'=e^{-v},\qquad  v'=\tfrac32 v_2-\tfrac34 v_1^2-u e^{2v};
\tag*{(\ref{8-1}) $\to$ (\ref{ds2})}
\\
u'=u_1+\sqrt{3}\,e^{v}(u_2+u_1v_1),\nonumber\\
v'=\tfrac32 v_2-\tfrac34 v_1^2-u e^{2v}
+\sqrt{3}\,v_1e^{-v}-\tfrac14 e^{-2v};   \tag*{(\ref{8-1}) $\to$
(\ref{ds2d})}
\\
u'=v,\qquad v'=-\tfrac32 v_2-v^2+u;   \tag*{(\ref{5}) $\to$
(\ref{8})}
\\
u'=v,\qquad v'=-\tfrac32 v_2- v^2+w,\qquad w_x=u,\qquad
w_t=u_2+uv;   \tag*{(\ref{5}) $\to$ (\ref{8-1})}
\\
u'=v_1,\qquad v'=-\tfrac32 v_3- v_1^2+u ;       \tag*{(\ref{5})
$\to$ (\ref{8-2})}
\\
u'=u+k,\qquad v'=v_1-\tfrac12v^2;  \tag*{(\ref{5}) $\to$
(\ref{5-1})}
\\
u'=v-k,\qquad v'=u_1-\tfrac32v_2 -\tfrac12(v^2+3k^2)+2kv;
\tag*{(\ref{5}) $\to$ (\ref{5-1-1})}
\\
u'=3u_1-\tfrac32v_1- \tfrac34(u^2+v^2),\nonumber\\
 v'=\tfrac94v_3-\tfrac94 u(u_2+2v_2)+\tfrac94vv_2-\tfrac92 u_1^2+\tfrac94 u^2(2u_1+v_1)\nonumber\\
\phantom{v'=}{}
-\tfrac94vv_1(2u+v)-\tfrac{9}{16}(u^2-v^2)^2;\label{11-15}\tag*{(\ref{5})
$\to$ (\ref{1d})}
\\
u' =-3u_2+\tfrac32 v_1,\nonumber\\
v' =\tfrac94 v_3+\tfrac92
u_1v_2-\tfrac{9}{16}(u_1^2-2u_2+2v_1)^2+\tfrac94(u_1^2+v_1)^2-\tfrac98
u_1^4; \label{11-16} \tag*{(\ref{5}) $\to$ (\ref{9-40})}
\\
u'=-3u_1+\frac32 v,\nonumber\\
v' =\tfrac94 v_2+\tfrac92
uv_1-\tfrac{9}{16}(u^2-2u_1+2v)^2+\tfrac94(u^2+v)^2-\tfrac98 u^4;
\label{11-17} \tag*{(\ref{5}) $\to$ (\ref{9-4})}
\\
u'=-3(v_1+2u_1)+\tfrac32 c_1g,\qquad g=u+v, \nonumber\\
v' =\tfrac92 v_3+\tfrac94 c_1(u_2-v_2)+18gv_2-9u_1^2+9v_1^2\nonumber \\
\phantom{v' =}{}
+9(c_1g+2g^2)u_1-\tfrac98c_1^2(u_1+g^2)+36g^2v_1+9g^4;
\label{11-18} \tag*{(\ref{5}) $\to$ (\ref{9-8})}
\\
u' =3v_1+u-3v^2, \nonumber\\
v' =-\tfrac92 v_3+\tfrac32
u_2+9vv_2-6u_1v+6(3v^2-u)v_1+3v^2(2u-3v^2); \tag*{(\ref{5}) $\to$
(\ref{11,1})}
\\
u'=-\frac{u_1+v_2}{\sqrt{u+v_1}}+c_1\sqrt{u+v_1}-\tfrac43v_1, \nonumber\\
 v'=\tfrac32\frac{u_3+v_4}{\sqrt{u+v_1}}-\tfrac{3}{4}c_1\frac{u_2+v_3}{\sqrt{u+v_1}}
 +\tfrac{8}{3}u\frac{u_1+v_2}{\sqrt{u+v_1}}
+\tfrac98\frac{(u_1+v_2)^3}{(u+v_1)^{5/2}}-\frac{(u_1+v_2)^2}{u+v_1}  \nonumber\\
\phantom{v'=}{}  -\tfrac38\frac{u1+v_2}{(u+v_1)^{3/2}}\big(6(u_2+v_3)-c_1(u_1+v_2)\big)+2(v_3+2u_2+c_1v_2)\nonumber\\
\phantom{v'=}{}
+\tfrac{16}{9}(u^2-v_1^2)-\tfrac83(v_2-u_1-c_1v_1)\sqrt{u+v_1}-\tfrac12c_1^2(u+2v_1);
\tag*{(\ref{5}) $\to$ (\ref{9-7})}
\\
u'=\sqrt{u+v_1}-k,\nonumber\\
v'=-\tfrac34\frac{u_2+v_3}{\sqrt{u+v_1}}+\tfrac38\frac{(u_1+v_2)^2}{(u+v_1)^{3/2}}+2k\sqrt{u+v_1}-v_1-\frac
u2 ; \label{11-24} \tag*{(\ref{5}) $\to$ (\ref{5-1-1-1})}
\\
u'=i\sqrt{6}\,u_1+u^2 +v,\qquad v'=\tfrac32( u_1^2-v_2)+
i\sqrt{6}\,uv_1-v^2;\tag*{(\ref{5}) $\to$ (\ref{s11a})}
\\
u' =3(u_1+v_1 -ku),\nonumber\\
v' =-\tfrac92v_3-\tfrac92 uu_2+9(u+k)v_2- \tfrac{27}{4}u_1^2-9v_1^2\nonumber\\
\phantom{v'=}{} +9u(u+k)u_1- 9(u^2+k^2)v_1-\tfrac94
u^2(u^2+2k^2);\tag*{(\ref{5}) $\to$ (\ref{s11b})}
\\
u'=-3u_2+\tfrac32v_1+\tfrac32 e^u(c_1+4\sqrt{c_2}\,u_1), \nonumber\\
 v'=\tfrac94v_3-\tfrac94\left(\tfrac12u_1^2-u_2+v_1+c_1e^u+2c_2e^{2u}\right)^2+\tfrac94c_1u_2e^u
 +\tfrac92u_1v_2+\tfrac94v_1^2\nonumber\\
\phantom{v'=}{} +\tfrac98u_1^2(u_1^2+4v_1)+\tfrac94e^u\big(c_1(5u_1^2+2v_1)-4\sqrt{c_2}\,( v_2+2u_1v_1+u_1^3)\big)+18c_2^2e^{4u}  \nonumber\\
\phantom{v'=}{}+\tfrac94e^{2u}\big(c_1(c_1-12\sqrt{c_2}\,u_1)+4c_2(3u_1^2+2v_1)\big)+18c_2e^{3u}(c_1-2\sqrt{c_2}\,u_1);
 \tag*{(\ref{5}) $\to$ (\ref{9-9})}
\\
u'=-3u_2-\tfrac32v_2-\tfrac34(u_1^2+v_1^2)- 2\sqrt{6c_2}\,(u_1+v_1)e^{u+v}-c_1e^{-2v}\nonumber\\
\phantom{u'=}{} -c_2e^{2(u+v)}+c_3e^{2(v-u)},\nonumber\\
 v'=\tfrac94v_4-\tfrac94 u_1u_3+\tfrac94v_3(2u_1+v_1)-\tfrac92u_2(u_2+u_1^2)-\tfrac{9}{16}(u_1^2-v_1^2)^2\nonumber\\
\phantom{v'=}{} +\tfrac94v_2(u_1^2+2u_1v_1-v_1^2)+4\sqrt{6c_2}\,v_1\big(2c_3e^{3v-u}-c_1e^{u-v}\big)+c\nonumber\\
\phantom{v'=}{}+3\sqrt{6c_2}\,\big(v_3-u_1u_2+v_2(2u_1+v_1)+2u_1v_1(u_1+v_1)\big)e^{u+v}\nonumber\\
\phantom{v'=}{}+6v_2\big(c_2e^{2(u+v)}+c_3e^{2(v-u)}-c_1e^{-2v}\big)+\tfrac32c_1(u_1-v_1)(u_1-3v_1)e^{-2v}\nonumber\\
\phantom{v'=}{}
-c_1^2e^{-4v}+4c_2c_3e^{4v}-\tfrac32c_2(u_1-v_1)^2e^{2(u+v)}+\tfrac32c_3(u_1-3v_1)^2e^{2(v-u)};
\tag*{(\ref{5}) $\to$ (\ref{2-1})}
\\
u'=3u_2-\tfrac32v_2-\tfrac34(u_1^2+v_1^2)-3ke^{-v}(u_1+2v_1)-3(c_1+2k^2)e^{-2v},\nonumber\\
 v'=\tfrac94v_4-\tfrac94 u_1u_3+\tfrac94v_3(v_1-2u_1)+\tfrac92u_2(u_1^2-u_2)-\tfrac{9}{16}(u_1^2-v_1^2)^2\nonumber\\
\phantom{v'=}{} +\tfrac94 v_2(u_1^2-2u_1v_1-v_1^2)-9c_1^2e^{-4v}+18kc_1e^{-3v}(u_1+2v_1)\nonumber\\
\phantom{v'=}{} +\tfrac92e^{-2v}\big(4k^2u_2-4c_1v_2-2k^2(u_1+v_1)^2+c_1(u_1+v_1)(u_1+3v_1)\big)\nonumber\\
\phantom{v'=}{}-\tfrac92k e^{-v}\big(u_3+2v_3-4u_2(u_1+v_1)+v_2(2v_1-3u_1)+u_1^2(u_1+2v_1)\big),\nonumber\\
\phantom{v'=}{}c_1=c^2-k^2;
 \tag*{(\ref{5}) $\to$ (\ref{2-3})}
\\
u'=-3u_2-\tfrac32v_2-\tfrac34(u_1^2+v_1^2)-3c_1^2e^{2(u+v)}-3c_1(3u_1+2v_1)e^{u+v}\nonumber\\
\phantom{u'=}{} +c_2e^{-u}-ke^{-2u},\qquad  k=3c_3^2, \nonumber\\
v'=\tfrac94(v_4-u_1u_3)+\tfrac94v_3(2u_1+v_1)-\tfrac92u_2(u_2+u_1^2)+\tfrac94v_2(u_1^2+2u_1v_1-v_1^2)\nonumber\\
\phantom{v'=}{} -\tfrac{9}{16}(u_1^2-v_1^2)^2-9c_1^4e^{4(u+v)}-18c_1^3(3u_1+2v_1)e^{3(u+v)}+6c_1^2c_2e^{u+2v} \nonumber\\
\phantom{v'=}{}+6kc_1^2e^{2u}  +2kc_2e^{-u-2v}+6kc_1(u_1-2v_1)e^{u-v}+\tfrac32c_2e^{-u}(u_2+2v_2+v_1^2)\nonumber\\
\phantom{v'=}{}-\tfrac92c_1e^{u+v}\big(u_3-2v_3+u_2(9u_1+4v_1)-v_2(u_1+ 2v_1)+4u_1^2(u_1+v_1)\big)\nonumber\\
\phantom{v'=}{}-\tfrac32ke^{-2v}\big(4v_2+u_1(4v_1-u_1)-3v_1^2\big)-k^2e^{-4v}+6c_1c_2e^v(u_1+2v_1);
\tag*{(\ref{5}) $\to$ (\ref{2-4})}
\\
u'=3u_2-\tfrac32v_2-\tfrac34(u_1^2+v_1^2)+3(c_1-c_2^2)e^{2(u+v)}+3c_2(u_1+2v_1)e^{u+v}, \nonumber\\
 v'=\tfrac94v_4-\tfrac94 u_1u_3+\tfrac94v_3(v_1-2u_1)+\tfrac92u_2(u_1^2-u_2)-\tfrac{9}{16}(u_1^2-v_1^2)^2\nonumber\\
\phantom{v'=}{} +\tfrac94 v_2(u_1^2-2u_1v_1-v_1^2)-9kc_2^2e^{4(u+v)}+18kc_2e^{3(u+v)}(u_1+2v_1)\nonumber\\
\phantom{v'=}{}+\tfrac92e^{2(u+v)}\big(2c_2^2u_2+4c_1v_2+c_1(u_1+3v_1)^2-c_2^2(u_1^2+4u_1v_1+5v_1^2)\big)\nonumber\\
\phantom{v'=}{}-\tfrac92c_2e^{u+v}\big(u_3+2v_3+u_2(u_1+4v_1)+v_2(2v_1-u_1)\big),
\qquad k=c_2^2-2c_1;\!\! \tag*{(\ref{5}) $\to$ (\ref{2-5})}
\\
u'=3u_2-\tfrac32v_2-\tfrac34(u_1^2+v_1^2)-3c_2^2e^{2(u+v)}+3c_2(u_1+2v_1)e^{u+v} \nonumber\\
\phantom{u'=}{}+2c_1c_2e^u+c_1e^{-v}(u_1+2v_1)- \tfrac23c_1^2e^{-2v}, \nonumber\\
 v'=\tfrac94v_4-\tfrac94 u_1u_3+\tfrac94v_3(v_1-2u_1)+\tfrac92u_2(u_1^2-u_2)-\tfrac{9}{16}(u_1^2-v_1^2)^2\nonumber\\
\phantom{v'=}{}+\tfrac94 v_2(u_1^2-2u_1v_1-v_1^2)-9c_2^4e^{4(u+v)}+18c_2^3e^{3(u+v)}(u_1+2v_1)\nonumber\\
\phantom{v'=}{}+12c_1c_2^3e^{3u+2v}+\tfrac92c_2^2e^{2(u+v)}(2u_2-u_1^2-4u_1v_1-5v_1^2)\nonumber\\
\phantom{v'=}{} -4c_1^2c_2^2e^{2u}+c_1^2e^{-2v}\big(2u_2-(u_1+v_1)^2\big)\nonumber\\
\phantom{v'=}{}-\tfrac92c_2e^{u+v}\big(u_3+2v_3+u_2(u_1+4v_1)+v_2(2v_1-u_1)\big)\nonumber\\
\phantom{v'=}{} -6c_1c_2^2e^{2u+v}(u_1+2v_1)-3c_1c_2e^u(3u_2+2v_2+2u_1v_1+3v_1^2)\nonumber\\
\phantom{v'=}{}+\tfrac32c_1e^{-v}\big(u_3+2v_3-4u_2(u_1+v_1)+v_2(2v_1-3u_1)+u_1^2(u_1+2v_1)\big);
\tag*{(\ref{5}) $\to$ (\ref{2-6})}
\\
u'=3u_2-\tfrac32v_2-\tfrac34(u_1^2+v_1^2)-\tfrac{15}{4} c_1^2e^{2(u+v)}+\tfrac92c_1(u_1+2v_1)e^{u+v} \nonumber\\
\phantom{u'=}{} +\tfrac52c_1c_2e^u+c_2e^{-v}(u_1+2v_1)- \tfrac{5}{12}c_2^2e^{-2v}, \nonumber\\
 v'=\tfrac94v_4-\tfrac94 u_1u_3+\tfrac94v_3(v_1-2u_1)+\tfrac92u_2(u_1^2-u_2)-\tfrac{9}{16}(u_1^2-v_1^2)^2\nonumber\\
\phantom{v'=}{} +\tfrac94 v_2(u_1^2-2u_1v_1-v_1^2)-\tfrac{81}{16}c_1^4\,e^{4(u+v)}+\tfrac{27}{4}c_1^3\,e^{3(u+v)}(u_1+2v_1)\nonumber\\
\phantom{v'=}{}+\tfrac{27}{4}c_1^3c_2e^{3u+2v}+\tfrac98c_1^2e^{2(u+v)}(18u_2+16v_2-5u_1^2-12u_1v_1-9v_1^2)\nonumber\\
\phantom{v'=}{} -\tfrac{27}{8}c_1^2c_2^2\,e^{2u}+\tfrac34c_1c_2^3\,e^{u-2v}-\tfrac94c_1c_2^2\,e^{u-v}(u_1+2v_1)\nonumber\\
\phantom{v'=}{}-\tfrac{27}{4}c_1e^{u+v}\big(u_3+2v_3+u_2(u_1+4v_1)+v_2(2v_1-u_1)\big)\nonumber\\
\phantom{v'=}{} -\tfrac34c_1c_2e^u(17u_2+14v_2+12u_1v_1+19v_1^2)+\tfrac12c_2^3e^{-3v}(u_1+2v_1)\nonumber\\
\phantom{v'=}{}+\tfrac18c_2^2e^{-2v}\big(16u_2+12v_2-11u_1^2-28u_1v_1-17v_1^2\big)-\tfrac{1}{16}c_2^4e^{-4v} \nonumber\\
\phantom{v'=}{}+\tfrac32c_2e^{-v}\big(u_3+2v_3-4u_2(u_1+v_1)+v_2(2v_1-3u_1)+u_1^2(u_1+2v_1)\big);
\tag*{(\ref{5}) $\to$ (\ref{2-8})}
\\
u' =3\frac{2g^3-u_2+2gv_1}{\sqrt{u_1-g^2}}+6g\sqrt{u_1-g^2}-6u_1-3v_1-3c(2g+c),\qquad g=u+v, \nonumber\\
v' =\tfrac92v_3+9(3g+c)u_2+9(2g+c)v_2-\tfrac94\frac{(2g^3-u_2+2gv_1)^2}{u_1-g^2} +9(2u_1+v_1)^2\nonumber\\
\phantom{v'=}{} -18u_1^2+18\sqrt{u_1-g^2}\big(v_2+2(2g+c)(v_1+g^2)\big)+36(2g+c)(u_1-g^2)^{3/2}\nonumber\\
\phantom{v'=}{}-9g^2u_1-18\big((g+c)^2+2g^2\big)v_1-36cg\big((g+c)^2+g^2\big)-45g^4;
\tag*{(\ref{5}) $\to$ (\ref{1d-6})}
\\
u'=v-2k,\qquad v'=u-\tfrac32v_1+\tfrac43k^3t;  \tag*{(\ref{5-1})
$\to$ (\ref{5-1-1})}
\\
\mbox{this substitution is invertible:}\nonumber\\
u'= \tfrac32u_1+v-\tfrac43k^3t,\qquad v'=u+2k ;
\tag*{(\ref{5-1-1}) $\to$ (\ref{5-1})}
\\
u'=-\tfrac12 v,\qquad v'=\sqrt{u+v_1};   \tag*{(\ref{5-1-1}) $\to$
(\ref{5-1-1-1})}
\\
u'=c_1\sqrt{2}\,e^{-v} ,\qquad v'=-v_1+\tfrac{2}{3}u\,e^{v} ;
\tag*{(\ref{1d}) $\to$  (\ref{11,1-0-0})}
\\
u'=ue^v,\qquad v'=v_1-c_1ue^v-2c_2;  \tag*{(\ref{1d}) $\to$
(\ref{e1})}
\\
u'=(u+\sqrt{-c}\,)e^v ,\qquad v'=v_1+(\sqrt{-c}-u)e^{v};
\tag*{(\ref{1d}) $\to$  (\ref{1d-5})}
\\
u'=u_1+2k\,e^{-v} ,\qquad v'=v_1-2c\,e^{-v};  \tag*{(\ref{1d})
$\to$  (\ref{2-3})}
\\
u' =u_1+ce^{u+v} ,\qquad v'=v_1-ke^{u+v},\nonumber\\
 \text{where  $c$ and $k$  are roots of } z^2-2c_2z+2c_1=0;
 \tag*{(\ref{1d}) $\to$  (\ref{2-5})}
\\
u'=u_1-\tfrac23c_1\,e^{-v} ,\qquad
v'=v_1-2c_2\,e^{u+v}+\tfrac23c_1\,e^{-v} ; \tag*{(\ref{1d}) $\to$
(\ref{2-6})}
\\
u'=u_1+2c_1\,e^{u+v} -\tfrac23\,c_2\,e^{-v},\qquad
v'=v_1-c_1\,e^{u+v} +\tfrac13\,c_2\,e^{-v};   \tag*{(\ref{1d})
$\to$  (\ref{2-8})}
\\
u'=2\sqrt{u_1-(u+v)^2} ,\qquad v'=2(u+v+c);   \tag*{(\ref{1d})
$\to$  (\ref{1d-6})}
\\
u'=-2v,\qquad v'=\tfrac23 u -2(v_1+v^2); \tag*{(\ref{9-4}) $\to$
(\ref{11,1})}
\\
\mbox{this substitution is invertible:}\nonumber\\
u'=\tfrac32(v-u_1)+\tfrac34 u^2,\qquad v'=-\tfrac 12 u;
\tag*{(\ref{11,1}) $\to$ (\ref{9-4})}
\\
u'=\tfrac 3 2 ( u_1-uv),\qquad v'=\tfrac 1 2 (u-v);
\tag*{(\ref{11,1}) $\to$ (\ref{1d})}
\\
u'=3(u+v)^2-3u_1+\tfrac{3}{16}c_1^2,\qquad v'=\tfrac{1}{4}c_1
-u-v;  \tag*{(\ref{11,1}) $\to$ (\ref{9-8})}
\\
u'=u,\qquad v'=v_1;  \tag*{(\ref{11,1}) $\to$ (\ref{11,2})}
\\
u'=\tfrac 4 3 u+\tfrac{3}{16}c_1^2,\qquad
v'=\tfrac{1}{4}c_1-\tfrac{2}{3} \sqrt{u+v_1};  \tag*{(\ref{11,1})
$\to$ (\ref{9-7})}
\\
u'=u,\qquad v'=\tfrac12\,v_1+\tfrac{1}{\sqrt{3}}\,c e^{-v};
\tag*{(\ref{11,1}) $\to$ (\ref{11,1-0})}
\\
u'=c_1\sqrt{2}\,u,\qquad v'=\tfrac 1 2 v_1-\tfrac 1 3
u\,e^{v}-\tfrac{\sqrt{2}}{2}\,c_1e^{-v};   \tag*{(\ref{11,1})
$\to$ (\ref{11,1-0-0})}
\\
u'=-u_1e^{v}-\tfrac13\,e^{2v}\left(2 u^2+c_1 \right),\qquad
v'=-\tfrac 1 2 v_1+\tfrac13\sqrt{-2c_1} \,e^{v} ;
\tag*{(\ref{11,1}) $\to$ (\ref{11,1-2})}
\\
u'=u_1e^{v}+\tfrac13\,e^{2v}\left(2 u^2+c_1 \right),\qquad
v'=-\tfrac 1 2 v_1+\tfrac23\,u\,e^{v};  \tag*{(\ref{11,1}) $\to$
(\ref{11,1-2})}
\\
u'=\tfrac32 e^{v}(u_1+2c_2u+c_1u^2\,e^{v}),\qquad v'=-\tfrac 1 2
v_1+\tfrac{1}{2}(c_1+1)u\,e^v+c_2;  \tag*{(\ref{11,1}) $\to$
(\ref{e1})}
\\
u'=\tfrac94(u+v)^2-\tfrac94\,u_1,\qquad v'=\tfrac94\,u ;
\tag*{(\ref{9-7}) $\to$ (\ref{9-8})}
\\
\mbox{this substitution is invertible:}\nonumber\\
u'=\tfrac{4}{9}\,v,\qquad
v'=\tfrac{2}{3}\,\sqrt{u+v_1}-\tfrac49\,v . \tag*{(\ref{9-8})
$\to$ (\ref{9-7})}
\end{gather}

The graph of the substitutions is very cumbersome, therefore we
show the most interesting subgraph only.

\begin{figure}[h]\centering
\unitlength=1mm \special{em:linewidth 0.6pt} \linethickness{0.6pt}
\begin{picture}(70.33,92.00)(60,56)
\font\plot=cmr10 at 9pt \plot \put(80.00,145.00){\circle{6.00}}
\put(80.30,145.00){\makebox(0,0)[cc]{{\ref{9}}}}
\put(65.00,130.00){\circle{6.00}}
\put(65.05,130.00){\makebox(0,0)[cc]{\ref{9-8}}}
\put(95.00,130.00){\circle{6.00}}
\put(95.07,130.00){\makebox(0,0)[cc]{\ref{9-7}}}
\put(78.0,143.0){\vector(-1,-1){11.0}}
\put(81.83,142.77){\vector(1,-1){11.0}}
\put(92.20,129.50){\vector(-1,0){24.4}}
\put(67.8,130.50){\vector(1,0){24.33}}
\put(80.00,113.67){\circle{6.00}}
\put(111.33,113.67){\circle{6.00}}
\put(111.45,113.67){\makebox(0,0)[cc]{\ref{5}}}
\put(80.00,113.67){\makebox(0,0)[cc]{\ref{8}}}
\put(82.0,115.80){\vector(1,1){11.67}}
\put(109.27,115.83){\vector(-1,1){12.23}}
\put(82.87,113.17){\vector(1,0){25.70}}
\put(108.5,114.33){\vector(-1,0){25.67}}
\put(82.47,112.40){\vector(3,-2){11.3}}
\put(95.67,102.67){\circle{6.00}}
\put(109.0,112.00){\vector(-3,-2){11.23}}
\put(95.67,83.33){\circle{6.00}}%
\put(95.67,86.13){\vector(0,1){13.80}}
\put(95.67,102.67){\makebox(0,0)[cc]{\ref{8-2}}}
\put(95.67,83.33){\makebox(0,0)[cc]{\ref{8-1}}}
\put(110.0,111.0){\vector(-1,-2){12.7}}
\put(80.70,111.0){\vector(1,-2){12.90}}
\put(78.67,111.20){\vector(-1,-3){13.65}}
\put(98.20,84.6){\vector(1,2){13.10}}
\put(64.00,67.67){\circle{6.00}}
\put(64.00,67.67){\makebox(0,0)[cc]{\ref{5-1-1}}}
\put(112.60,111.0){\vector(1,-3){13.55}}
\put(127.00,67.67){\circle{6.00}}
\put(127.00,67.67){\makebox(0,0)[cc]{\ref{5-1}}}
\put(124.3,68.33){\vector(-1,0){57.43}}
\put(66.83,67.33){\vector(1,0){57.40}}
\put(111.67,83.00){\circle{6.00}}
\put(111.7,83.00){\makebox(0,0)[cc]{\ref{ds2d}}}
\put(79.67,83.00){\circle{6.00}}
\put(92.90,83.00){\vector(-1,0){10.50}}
\put(98.40,83.00){\vector(1,0){10.50}}
\put(79.7,83.00){\makebox(0,0)[cc]{\ref{ds2}}} \put(61,58){{{\bf
Fig. 1.} A subgraph of the dif\/ferential substitutions.}}
\end{picture}
\end{figure}

\noindent {\bf Comments.} (1) System (\ref{9}) coincides with
(\ref{s1}) and (\ref{8}) coincides with (\ref{s2}).

(2) It was simpler to obtain some systems from (\ref{5}) than
(\ref{8}) or vice versa. These systems are connected by the second
order invertible substitution, see (\ref{5}) $\to$ (\ref{8}) and
(\ref{8}) $\to$ (\ref{5}). Hence, each system obtained from
(\ref{5}) can be obtained from
 (\ref{8})  and vice versa. (3) Fifteen systems
(\ref{1})--(\ref{11,2}), (\ref{9-7}), (\ref{9-9}),
(\ref{2-3})--(\ref{2-6}), (\ref{2-8}) and (\ref{1d-6}) can be
obtained from both  (\ref{9}) and (\ref{5}) by the presented
dif\/ferential substitutions. Twelve systems (\ref{9-0}),
(\ref{9-2}), (\ref{9-5}), (\ref{11,1-0}), (\ref{11,1-0-0}),
(\ref{11,1-1})--(\ref{1d-5}),
 (\ref{2-2}), (\ref{2-7})  and (\ref{2-9}) can be obtained from  (\ref{9}). The remaining eleven systems (\ref{8})--(\ref{8-2}),
 (\ref{5-1}), (\ref{5-1-1}),
(\ref{5-1-1-1})--(\ref{s11b}), (\ref{ds2}), (\ref{ds2d}) and
(\ref{2-1}) can be obtained from  (\ref{5}) or (\ref{8}).

\begin{remark}
  As the systems  (\ref{9}) and (\ref{8}) have the Lax  representations, then all  systems from the list have the Lax  representations in
a generalized meaning (see Section~\ref{sec5}).
\end{remark}

\begin{remark}
Some of the presented substitutions are superpositions of lower
order substitutions, other substitutions are irreducible.
\end{remark}

\begin{remark}
System (\ref{9}) admits the f\/irst and second order substitutions
and does not  admit the third and fourth order substitutions.
Probably it does not  admit any higher order substitutions,
either. Systems (\ref{8-1}) and (\ref{5}) admit the substitutions
from the f\/irst till fourth orders. We do not present higher
order substitutions for  (\ref{8-1}) because simpler substitutions
exist for  (\ref{5}). Fifth and higher order substitutions for
systems (\ref{8-1}) and  (\ref{5}) have not been computed because
the computations are extremely cumbersome.
\end{remark}

\begin{remark}
There are some additional dif\/ferential substitutions under the
constraints for constants in the systems. For example, there are
substitutions $(\ref{9})\to(\ref{2-1})$ for $c_1=0$  and
$(\ref{5})\to(\ref{2-2})$ for $c_3=0$ or $(\ref{5})\to(\ref{2-9})$
for $c_2=0$ and so on. These substitutions are not so important
and we do not present them here.
\end{remark}

\begin{remark}
Unexpectedly, the well known systems (\ref{9}) and (\ref{8}) are
implicitly connected as it is shown in Fig.~1.
\end{remark}

\section{Examples of zero curvature representations}\label{sec5}

The IST method for nonlinear equations with two independent
variables is based on investigation of a linear overdetermined
system
\begin{gather}\label{LA}
L(\bsy u,\lambda,\p_x )\psi =0,\qquad \psi_t=A(\bsy u,\lambda,\p_x
)\psi ,
\end{gather}
where $L$ and $A$ are ordinary linear operators, $\bsy u$ is a
smooth (vector) function satisfying a~nonlinear partial
dif\/ferential equation $E(\bsy u)=0$ and $\lambda$ is a
parameter. The operators~$L$, $A$ may be both scalar and matrix.
The operators~$L$,~$A$ and $E$ may be also pseudodif\/ferential or
integro-dif\/ferential.  The compatibility condition of system
(\ref{LA}) reads
\begin{gather*}
\left(\frac{\p L}{\p t}+LA\right)\psi \cond{L\psi =0}=0.
\end{gather*}
There are two ways to satisfy this condition. The f\/irst operator
condition was introduced by P.D.~Lax~\cite{PL}:
\begin{gather}\label{L-A}
\frac{\p L}{\p t}+LA=AL, \qquad \text{or}\qquad \frac{\p L}{\p
t}=[A,L].
\end{gather}
The second more general operator condition was introduced in
\cite{Man}, see also~\cite{Zah}:
\begin{gather}\label{LAB}
\frac{\p L}{\p t}+LA=BL.
\end{gather}
If an equation $E(\bsy u)=0$ is equivalent to equation
(\ref{L-A}), then  (\ref{L-A}) is said to be the Lax
representation of the equation $E(\bsy u)=0$. The pair of
operators $(L,A)$ is said to be the $(L,A)$-pair or the Lax pair.

If an equation $E(\bsy u)=0$ is equivalent to equation
(\ref{LAB}), then  (\ref{LAB}) is said to be the $(L,A,B)$
representation of the equation $E(\bsy u)=0$ or the triad
representation.

In all cases, operator $L$ must essentially depend on the
parameter $\lambda$. This parameter cannot be removed by a gauge
transformation $L\to f^{-1}Lf$ with some smooth function $f$, in
particular.

If system (\ref{LA}) is dif\/ferential, then the  standard
substitution $\psi =\Psi_1$, $\psi_{x}=\Psi_2$ and so on, provides
the following f\/irst order system:
\begin{gather}\label{UV}
\Psi_x=U\Psi,\qquad \Psi_t=V\Psi,
\end{gather}
where $U$ and $V$ are square matrices depending on $\bsy u$ and
$\lambda$. The compatibility condition of linear system (\ref{UV})
reads
\begin{gather}\label{zer}
U_t-V_x+[U,V]=0
\end{gather}
if $E(\bsy u)=0$. Usually a stronger condition is required:
(\ref{zer}) is valid if\/f $E(\bsy u)=0$. In this case equation
(\ref{zer}) is said to be the
 zero curvature representation. For an evolutionary system $\bsy u_t=\bsy K(\bsy u,\bsy u_x,\dots,\bsy u_n),\ \bsy u=\{u^\alpha\}$ the
matrix $U$ usually depend on $\bsy u$ only, but it may depend on
$\bsy u$, $\bsy u_x$, $\bsy u_{xx}$, and so on. Let us consider
the general case.

If some smooth functions $F=F(\bsy u,\bsy u_x,\dots,\bsy u_r)$ and
$\Phi=\Phi(\bsy u,\bsy u_x,\dots,\bsy u_p)$ satisfy the condition
\begin{gather}\label{usl}
\big(\p_tF+\Phi\big)\cond{\bsy u_t=\bsy K}=0,
\end{gather}
then one obtains
\[
\Phi\cond{\bsy u_t=\bsy K}=\Phi,\ \ \p_tF\cond{\bsy u_t=\bsy K}=
\frac{\p F}{\p u^\alpha_i }\p_tu^\alpha_i\cond{\bsy u_t=\bsy K} =
\frac{\p F}{\p u^\alpha_i }\p_x^iu^\alpha_t\cond{\bsy u_t=\bsy K}=
\frac{\p F}{\p u^\alpha_i }D_x^iK^\alpha,
\]
where the summation over $i=0,\dots,r$ and $\alpha=1,\dots,m$ is
implied. This implies
\[
\Phi=- \frac{\p F}{\p u^\alpha_i }D_x^iK^\alpha
\]
according to (\ref{usl}). Using this result one  obtains the
following identity:
\begin{gather}\label{res}
\p_tF+\Phi\equiv \frac{\p F}{\p u^\alpha_i }D_x^i(u_t^\alpha
-K^\alpha ),\qquad \forall\, \bsy u
\end{gather}
for any $F$ and $\Phi$ satisfying (\ref{usl}).

Let us apply identity (\ref{res})  to equation (\ref{zer}). If the
matrix $U$ depends on $\bsy u$ only then
\begin{gather}\label{rU}
U_t-V_x+[U,V]= \frac{\p U}{\p u^\alpha }(u_t^\alpha
-K^\alpha),\qquad \forall \, \bsy u.
\end{gather}
It is obvious now that equation (\ref{zer}) is equivalent to $\bsy
u_t=\bsy K$ if\/f the  matrices $\p U/\p u^\alpha$, $\alpha
=1,\dots,m $ are linearly independent. Suppose now that the matrix
$U$ depends on $\bsy u$ and $\bsy u_x$ then one obtains
\begin{gather}\label{rU1}
U_t-V_x+[U,V]= \frac{\p U}{\p u^\alpha }(u_t^\alpha -K^\alpha)+
\frac{\p U}{\p u^\alpha_x }D_x(u_t^\alpha -K^\alpha ),\qquad
\forall\, \bsy u.
\end{gather}
If the matrices $\p U/\p u^\alpha$, $\p U/\p u^\beta _x$, $\alpha,
\beta=1,\dots,m$ are linearly independent, then equation
(\ref{zer}) is equivalent to $\bsy u_t=\bsy K$ again.   Otherwise,
equation (\ref{zer}) would be equivalent to some dif\/ferential
consequence of the system $\bsy u_t=\bsy K$ that is a more general
system than the original one. In this case we call equation
(\ref{zer}) the {\it generalized} zero curvature representation.

It is well known that equations (\ref{UV}) and (\ref{zer}) are
covariant under the following transformation
\begin{gather}\label{gau}
\bar \Psi =S^{-1}\Psi,\qquad \bar U=S^{-1}(US-S_x),\qquad \bar
V=S^{-1}(VS-S_t),
\end{gather}
where $S$ is any non-degenerate matrix. This transformation is
called a gauge one. Any gauge transformation is invertible and
preserves compatibility of system (\ref{UV}).

Two $(L,A)$-pairs were proposed for system (\ref{s1}) in
\cite{DS1}. One of these $(L,A)$-pairs coincides with the
$(L,A)$-pair that was presented in \cite{HS}. The $L$-operator of
the common $(L,A)$-pair takes the form $L=(\p_x^2+f)(\p_x^2-g)$,
where
\[
f=\tfrac16(u \sqrt{2}-2v),\qquad g=\tfrac16(u\sqrt{2}+2v).
\]
 The temporal Lax equation reads  $\psi_t=A\psi$, where $A$ is a fractional degree of $L$.
The spatial Lax equation $L\psi =\lambda^2 \psi$ can be
transformed into the system  $(\p_x^2-g)\psi =\lambda \vph$,
$(\p_x^2+f)\vph=\lambda \psi $ and then into the normal form
(\ref{UV}), were
\begin{gather}
U=\begin{pmatrix} 0 & 1 & 0 & 0 \\ (u\sqrt{2}-2v)/6 & 0& \lambda &
0\\ 0&0&0&1\\ \lambda & 0 &-(u\sqrt{2}+2v)/6 &0
\end{pmatrix},\nonumber\\
V=\begin{pmatrix}
(u_1\sqrt{2}+v_1)/6 & -(u_1\sqrt{2}+v_1)/3 & 0 &-2\lambda \\[1mm] f_1+f_2 & -(u_1\sqrt{2}+v_1)/6 & \lambda v/3 & 0\\[1mm]
0 & -2\lambda & -(u_1\sqrt{2}-v_1)/6 & (u_1\sqrt{2}-v_1)/3\\[1mm] \lambda v/3 & 0 & f_1-f_2 & (u_1\sqrt{2}-v_1)/6
\end{pmatrix}.\label{U1}
\end{gather}
Here $f_1$ and $f_2$ take the following form:
\[
f_1 =\tfrac{1}{18}(3v_2-2u^2+2v^2)-2\lambda^2,\qquad f_2=
\tfrac{\sqrt{2}}{18}(3u_2+uv).
\]
Matrices (\ref{U1}) realize the zero curvature representation of
system (\ref{s1}).

System (\ref{s2}) also has two Lax  representations  (see
\cite{DS1}). Using the simpler $L$-operator, we have found,
similarly to the previous case, the following matrices that
realize the zero  curvature representation of  system (\ref{s2}):
\begin{gather}
\t U=\begin{pmatrix} 0 & 1 & 0 & 0 \\ -v/3-\lambda  & 0& 1 & 0 \\
0&0&0&1\\ u/9 & 0 &\lambda -v/3 &0
\end{pmatrix},\nonumber\\
\t V=\begin{pmatrix}
v_1/6 & 2\lambda -v/3 & 0 &-2 \\[1mm] h-\lambda v/3 & -v_1/6 & v/3 & 0\\[1mm]
u_1/9 & -2u/9 & v_1/6 & -2\lambda -v/3\\[1mm] uv/27+u_2/9 & -u_1/9 & h+\lambda v/3 & -v_1/6
\end{pmatrix},\label{U2}
\end{gather}
where
\[
h=\tfrac{1}{6}v_2+\tfrac{1}{9}(v^2-2u)-2\lambda^2.
\]

Let us consider an admissible dif\/ferential substitution $u=f(\t
u_i,\t v_j)$, $v=g(\t u_i,\t v_j)$  of system~(\ref{sys0}).
Substituting $u$ and $v$ in the matrices $U(u,v)$ and $V(u_i,v_j)$
one obtains
\[
\hat U(\t u_i,\t v_j)=U(f,g),\qquad \hat V(\t u_i,\t
v_j)=V(D_x^kf,D_x^lg).
\]
As $\hat U$ depends on derivatives of $\t u$ or $\t v$, then one
has a generalized zero curvature representation.

To obtain an ordinary zero curvature representation one can try to
remove higher order derivatives from the matrix $\hat U$ using the
gauge transformation (\ref{gau}). But this is not always possible
(see example B below).

{\bf A.} Performing the substitution (\ref{9}) $\to$ (\ref{1})
into the matrix $U$ from (\ref{U1}) one obtains
\[
\hat U=\begin{pmatrix} 0 & 1 & 0 & 0 \\ (u_2-v_2)/2+h_1 & 0&
\lambda & 0\\ 0&0&0&1\\ \lambda & 0 &-(u_2+v_2)/2+h_2 &0
\end{pmatrix},
\]
where
\[
h_1=\tfrac14(u_1-v_1)^2,\qquad h_2=\tfrac14(u_1+v_1)^2.
\]
One can easily verify that matrices $A=\hat U_{u_2}$ and $B=\hat
U_{v_2}$ are commutative, hence the system
$S_x=(Au_{xx}+Bv_{xx})S$ has the following solution
$S=\exp(Au_{x}+Bv_{x})$. The matrix $U_1=\bar{\hat U}$ evaluated
according to (\ref{gau}) takes the following form
\[
U_1=\frac12\begin{pmatrix} u_x-v_x & 2 & 0 & 0 \\ 0 & v_x-u_x &
2\lambda & 0\\ 0&0& -u_x-v_x &2 \\ 2\lambda & 0 & 0 &u_x+v_x
\end{pmatrix}.
\]
Now another gauge transformation is possible with the following
diagonal matrix:
\[
S_1=\exp\left(\int\text{diag\,}(U_1)dx\right),
\]
where $\text{diag}\,(U_1)$ is the main diagonal of $U_1$. This
gauge  transformation provides the following matrix
\[
U_2=\begin{pmatrix} 0 & e^{v-u} & 0 & 0 \\ 0 & 0 & \lambda e^{-v}
& 0\\ 0&0& 0 &e^{u+v} \\ \lambda e^{-v} & 0 & 0 & 0
\end{pmatrix}.
\]
A corresponding $V$-matrix can be obtained by solving equation
(\ref{zer}) directly or by the previous twofold gauge
transformation. This matrix takes the following form
\[
V_2=\begin{pmatrix}
0 & e^{v-u}f_1 & \lambda e^{-u}(u_1+v_1) & -2\lambda e^v \\[1mm]
-2\lambda^2e^{u-v}  & 0 & \dfrac{\lambda}{4} e^{-v}(2v_2-3u_1^2+v_1^2) & \lambda e^u(u_1-v_1) \\[1mm]
\lambda e^u(v_1-u_1)& -2\lambda e^v & 0 &e^{u+v}f_2 \\[1mm]
\dfrac{\lambda}{4} e^{-v}(2v_2-3u_1^2+v_1^2) & -\lambda
e^{-u}(u_1+v_1) & -2\lambda^2e^{-u-v} & 0
\end{pmatrix},
\]
where
\[
f_1=-u_2-\tfrac{1}{2}v_2+u_1v_1+\tfrac{1}{4}(u_1^2+v_1^2),\qquad
f_2=u_2-\tfrac{1}{2}v_2-u_1v_1+\tfrac{1}{4}(u_1^2+v_1^2).
\]
The matrices $U_2$ and $V_2$  realize the zero  curvature
representation of  system (\ref{1}).

{\bf B.} Substitution (\ref{9}) $\to$ (\ref{9-5}) reduces matrix
$U$ from (\ref{U1}) to the following form
\[
\hat U=\begin{pmatrix} 0 & 1 & 0 & 0 \\ (k-v)/3+R & 0& \lambda &
0\\ 0&0&0&1\\ \lambda & 0 &(k-v)/3-R &0
\end{pmatrix},
\]
where $R=\sqrt{2(u_1+v^2)}/6$. It is obvious that one cannot
remove $u_1$ from $\hat U$ by a gauge transformation. It is clear
from the structure of the matrix $\hat U$ that
\begin{gather*}
U_t-V_x+[U,V] =A\left(kv_1+\tfrac12u_2-\tfrac12v_3-v_t\right)\\
\qquad{}+BD_x\left(u_3-\tfrac{3}{4}\,\frac{(2\,v\,v_1+ u_2)^2}{v^2
+u_1}+3v\,v_2 +\tfrac{3}{2}v_1^2 +\tfrac{2}{3}v^3- k(2
v^2+u_1)-u_t\right),
\end{gather*}
where $A$ and $B$ are some linearly independent matrices. Thus,
this zero  curvature representation for  system (\ref{9-5}) is
generalized. Of course, one may introduce here the new variable
$u'=u_1$ to obtain an ordinary zero  curvature representation. But
we do not know if it is always possible.

{\bf C.} Performing the substitution (\ref{8}) $\to$ (\ref{5})
into the matrix $U$ from (\ref{U2}) one obtains
\[
\hat{\t U}=\begin{pmatrix} 0 & 1 & 0 & 0 \\ -u/3-\lambda  & 0& 1 &
0 \\ 0&0&0&1\\ u_2/6+(u^2+v)/9 & 0 &\lambda -u/3 &0
\end{pmatrix}.
\]
The f\/irst gauge transformation is performed using
$S_1=\exp(u_1(\p \hat{\t U}/\p u_2))$:
\[
S_1=\begin{pmatrix} 1 & 0 & 0 & 0 \\ 0  & 1& 0 & 0 \\ 0 &0&1&0 \\
u_1/6 & 0 &0 & 1
\end{pmatrix}.
\]
The transformed $U$-matrix is
\[
\t U_1=\begin{pmatrix} 0 & 1 & 0 & 0 \\ -u/3-\lambda  & 0& 1 & 0
\\ u_1/6 &0 & 0 & 1\\ (u^2+v)/9 & -u_1/6 &\lambda -u/3 &0
\end{pmatrix}.
\]
The second gauge transformation is performed using
$S_2=\exp(u\,(\p\t U_1/\p u_1))$:
\[
S_2=\begin{pmatrix} 1 & 0 & 0 & 0 \\ 0  & 1& 0 & 0 \\ u/6 &0&1&0
\\ 0 & -u/6 &0 & 1
\end{pmatrix}.
\]
The result of the twofold gauge transformation is
\begin{gather*}
 \t U_2=\begin{pmatrix}
0 & 1 & 0 & 0 \\ -u/6-\lambda  & 0& 1 & 0 \\ 0 & -u/3 & 0 & 1\\
u^2/36+v/9 & 0 &\lambda -u/6 &0
\end{pmatrix}, \\
 \t V_2=\begin{pmatrix}
-u_1/6 & 2\lambda & 0 & -2 \\[1mm] f_3-\lambda u/3 & -u_1/6 & u/3 & 0 \\[1mm]
(uu_1-v_1)/18-\lambda u_1/3 & -u_2/3-u^2/6-2v/9 & u_1/6 & -2\lambda  \\[1mm]
f_4 & (v_1-uu_1)/18-\lambda u_1/3 & f_3+\lambda u/3& u_1/6
\end{pmatrix},
\end{gather*}
where
\[
f_3=-\tfrac16 u_2-\tfrac{1}{18}u^2-\tfrac29v-2\lambda ^2,\qquad
f_4=\tfrac{1}{18}(uu_2-v_2+u_1^2)+\tfrac{1}{108}u^3-\tfrac{1}{27}uv-\tfrac23
u\lambda^2.
\]
Matrices $\t U_2$ and $\t V_2$ realize the zero curvature
representation of system (\ref{5}).

\section{Conclusion}
The examples in Section \ref{sec5} illustrate the fact that some
systems possess ordinary  zero curvature representation while
others possess generalized  zero curvature representation. All
these  representations are obtained from the Drinfeld--Sokolov
$L$, $A$ operators by using corresponding dif\/ferential
substitutions listed in Section~\ref{sec4}. Matrices $U$ and $V$
that realize all zero curvature representations have the size
$4\times 4$. Thus, the two-f\/ield evolutionary systems presented
above are integrable in principle by the inverse spectral
transform method. But the fact is that the inverse scattering
problem for dif\/ferential equations with order more than two is
extremely dif\/f\/icult. That is why other methods for solution of
equations may be useful~\cite{DD}. They may be B\"acklund
transformations \cite{Mi}, Darboux transformations \cite{MS, Gu},
Hirota method \cite{Hi} or numeric simulating (see \cite{Zen}, for
example).

\subsection*{Acknowledgments}
We are grateful to Professor V.V. Sokolov for helpful discussions.
This work was supported by Fe\-de\-ral Agency for Education of
Russian Federation, project \# 1.5.07.

\pdfbookmark[1]{References}{ref}
\LastPageEnding

\end{document}